\documentclass[bibyear]{aa} 

\newcommand{\pasiphae}{PASIPHAE}

\usepackage{natbib}

\usepackage[labelfont=bf]{caption} 
\usepackage{graphicx}
\usepackage{txfonts}
\usepackage{subcaption}
\usepackage{mathtools}

\begin{document} 

   \title{Local measurements of the mean interstellar polarization at high Galactic latitudes}

   \subtitle{}

   \author{R. Skalidis,\inst{1,}\inst{2}\fnmsep\thanks{E-mail: rskalidis@physics.uoc.gr}
          G. V. Panopoulou\inst{3}, K. Tassis\inst{1,}\inst{2},
          V. Pavlidou\inst{1,}\inst{2}, D. Blinov\inst{1,}\inst{2,}\inst{4},
          I. Komis\inst{1}, I. Liodakis\inst{5}
          }

   \institute{Department of Physics and Institute for Theoretical and Computational Physics, University of Crete, GR-71003 Heraklion
    Greece
    \and
   Foundation for Research and Technology – Hellas, IESL, Voutes, GR-7110 
   Heraklion, Greece
   \and Cahill Center for Astronomy and Astrophysics,
   California Institute of Technology, 1200 E California Blvd, MC 249-17,
   Pasadena CA, 91125, USA \and 
   Astronomical Institute, St. Petersburg State University, Universitetsky pr. 
   28, Petrodvoretz, 198504 St. Petersburg, Russia \and
   KIPAC, Stanford University, 452 Lomita Mall, Stanford, CA 94305, USA}

   \date{}

 
  \abstract{
    Very little information exists concerning the properties of the interstellar medium (ISM)-induced starlight polarization at high Galactic latitudes. Future optopolarimetric surveys promise to fill this gap. We conduct a small-scale pathfinding survey designed to identify the average polarization properties of the diffuse ISM locally, at regions with the lowest dust content.
We perform deep optopolarimetric surveys within three $\sim 15' \times 15'$ regions located at $b > 48^\circ$ using the RoboPol polarimeter. The observed samples of stars are photometrically complete to $\sim$16 mag in the R-band. The selected regions exhibit low total reddening compared to the majority of high-latitude sightlines. We measure the level of systematic uncertainty for all observing epochs and find it to be 0.1\% in fractional linear polarization, $p$.
The majority of individual stellar measurements have low signal-to-noise ratios. However, our survey strategy enables us to locate the mean fractional linear polarization $p_{mean}$ in each of the three regions. The region with lowest dust content yields $p_{mean}=(0.054 \pm 0.038) \%$,  not significantly different from zero. We find significant detections for the remaining two regions of: $p_{mean}=(0.113 \pm 0.036) \%$ and $p_{mean}=(0.208 \pm 0.044) \%$. Using a Bayesian approach, we provide upper limits on the intrinsic spread of the small-scale distributions of $q$ and $u$.
At the detected $p_{mean}$ levels, the determination of the systematic uncertainty is critical for the reliability of the measurements. We verify the significance of our detections with statistical tests, accounting for all sources of uncertainty. Using publicly available HI emission data, we identify the velocity components that most likely account for the observed $p_{mean}$ and find their morphologies to be misaligned with the orientation of the mean polarization at a spatial resolution of 10$\arcmin$. We find indications that the standard upper envelope of $p$ with reddening underestimates the maximum $p$ at very low E(B-V) ($\leq 0.01$ mag).}

\titlerunning{Local measurements of $p_{mean}$ at
high galactic latitudes}
\authorrunning{R. Skalidis et al.}
   \maketitle

%

\section{Introduction}
  
  Galactic dust is ubiquitous throughout the sky \citep[e.g.,][]{dust_map} and interacts with the large-scale Galactic magnetic field. Asymmetric dust grains tend to orient their short axis along the magnetic field lines. The most plausible mechanism of alignment is given by radiative alignment torque theory (RAT) \citep[for a recent review on grain alignment, see][]{andersson}. As a result of this alignment, the dust thermal emission is polarized perpendicular to this axis \citep{cudlip,stein} at far-infrared (FIR). On the other hand, starlight that passes through a dusty region suffers dichroic extinction; this results in the starlight becoming polarized parallel to the field lines \citep{hiltner2,hall1949,davisgreenstein}. Therefore, the plane-of-sky component of the magnetic field can be traced through the polarization of starlight caused by dust. 
  
  The fractional linear polarization, \textit{p}, is related to the dust column density, and therefore to stellar reddening, E(B-V). Observed values of $p$ are bounded by the empirical upper limit $p_{max}\rm =9(\%)E(B-V)$ \citep{hiltner1956,serkowski1975}. The majority of existing optical polarization observations have been driven by star formation studies, and consequently are agglomerated near the Galactic plane \citep[e.g.,][]{heiles}, that is, at regions with high E(B-V). 
There have been, however, a number of works targeting stars that have low E(B-V), either because they lie in our local neighborhood \citep{piirola,tinbergen,leroy,bailey}, or because they are located at high Galactic latitudes \citep[e.g.,][and references therein]{appenzeller,berdyugin2014}. The samples of stars in these surveys have been selected on the basis of distance, and consist entirely of bright stars (${\rm V}<13 \, {\rm mag}$ for the deepest sample of \cite{berdyugin2014}, which extends out to $\leq 600$ pc from the Sun). These works have measured the level of interstellar polarization towards individual stars that are spread out over several thousand square degrees. Though informative, these sparsely sampled (in all three dimensions) datasets form an incomplete picture of interstellar polarization at low extinctions. 

     \begin{figure*}
    \centering
    \includegraphics[width=\textwidth]{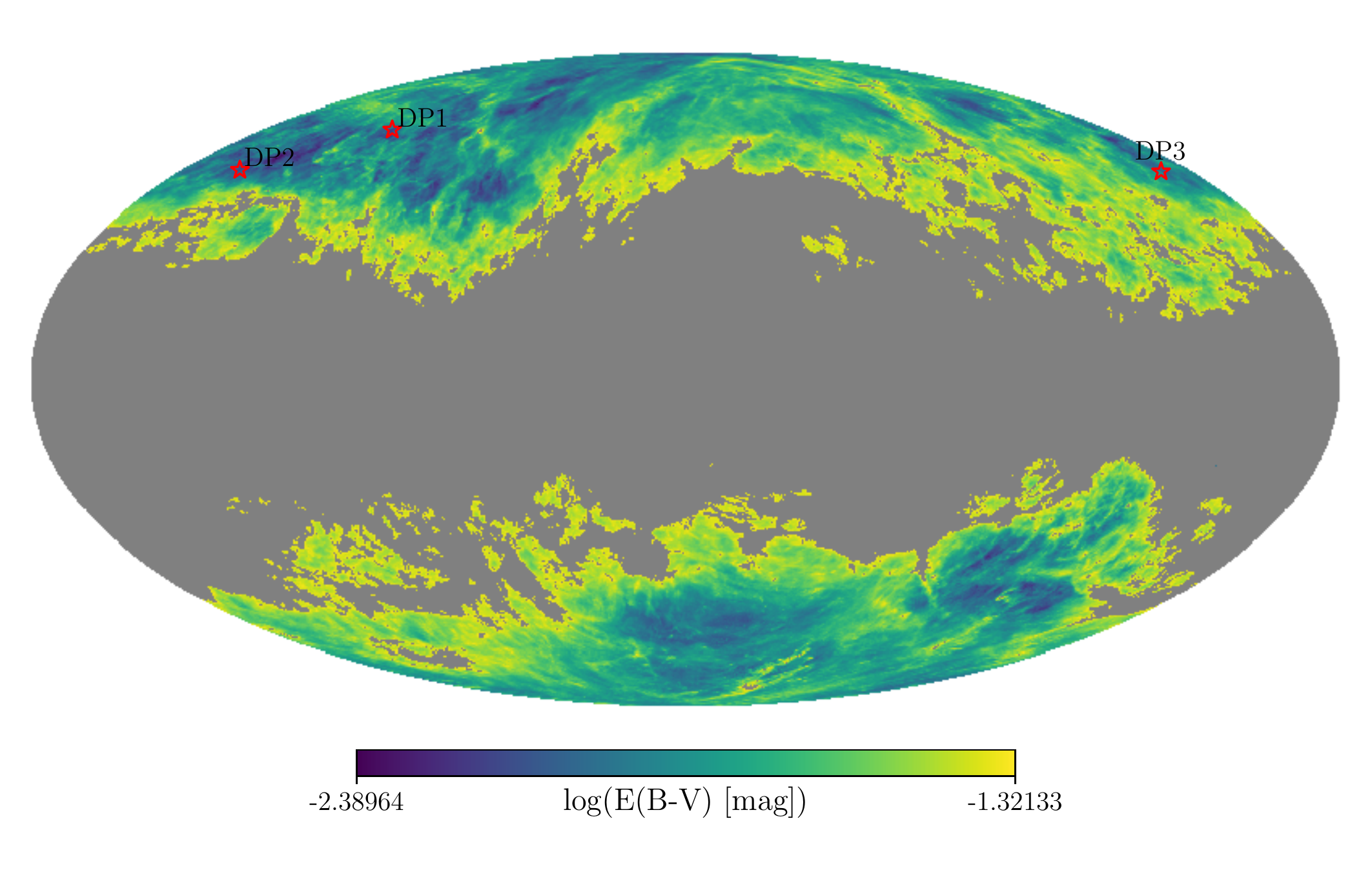}
    \caption{Mollweide projection of the E(B-V) map of \cite{lenz} (centered at l,b = [0,0]; grid-line spacing is 30$^\circ$). The red stars mark our target regions. Gray areas are not included in the map. }
    \label{fig:lenz_map}
  \end{figure*}

  
  
  The interest in understanding interstellar medium (ISM) polarization in this low dust-column regime is multifaceted. There is much to be gained in terms of understanding of the Galactic magnetic field and its effect on the diffuse ISM. In addition, this regime can offer new insights regarding the micro-physical interaction of dust with the magnetic field. A third and largely sought-after reward relates to the role of Galactic dust as a foreground to the cosmic microwave background (CMB). 
  
  Polarized thermal dust emission from our Galaxy is a major obstacle in the search for the primordial B-mode signal in the polarization of the CMB \citep{bicep}. This signal is predicted to arise from the effect of gravitational waves on the last scattering surface after the inflationary epoch \citep{seljak,seljak_zalda,kamionkowski1,zaldarriaga1,kamionkowski2}.

  \begin{figure}
   \centering
     \includegraphics[width=\hsize]{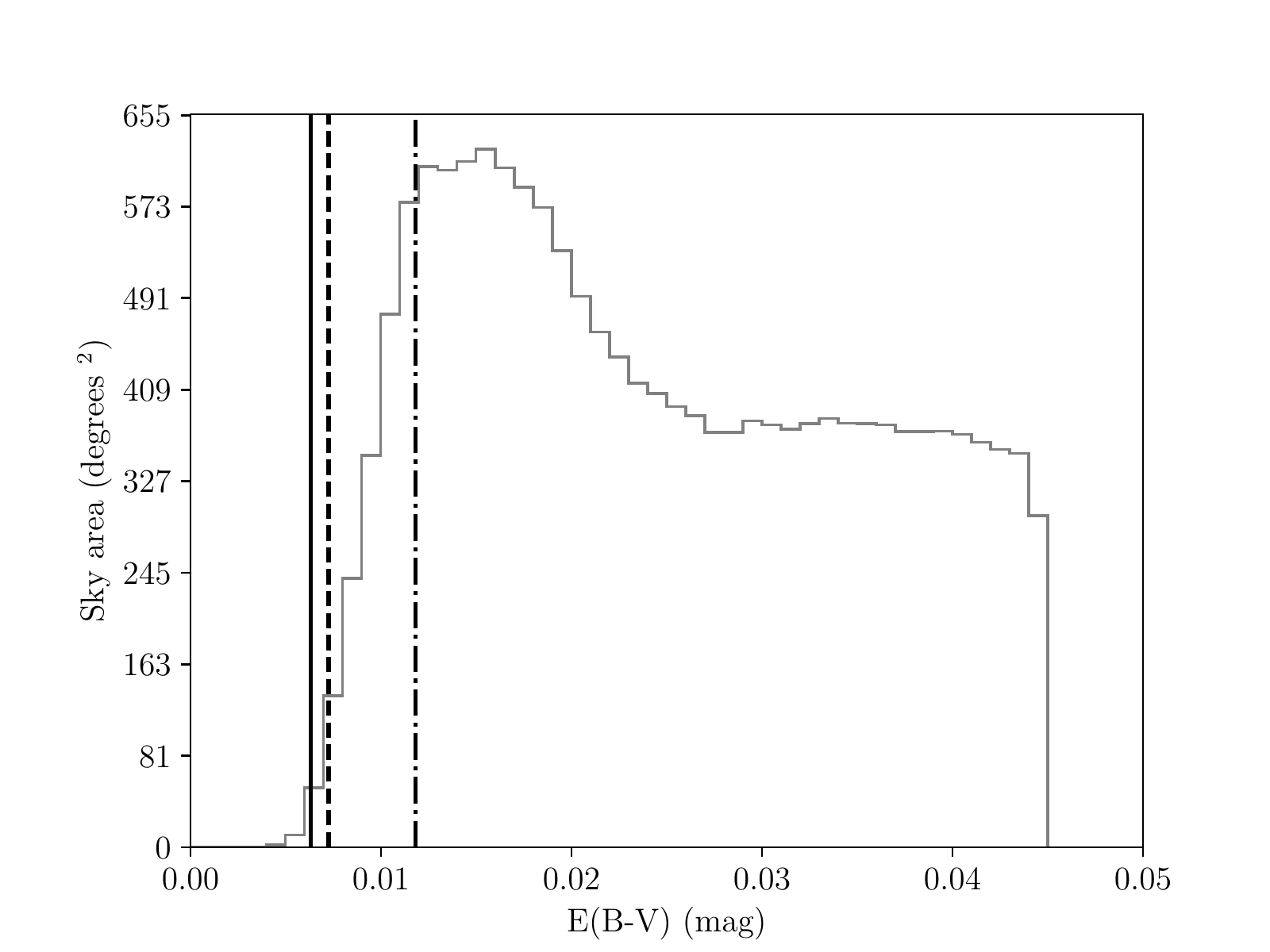}
     \caption{Distribution of E(B-V) from the map of 
              \cite{lenz}. Vertical lines correspond to the E(B-V) of the target regions DP1, DP2, and DP3. The truncation at 0.045 mag corresponds to the mask placed by Lenz et al. at hydrogen column densities $N_{HI} > 4\times 10^{20} \textrm{cm}^{-2}$.}
      \label{fig:dark_patch_distrib}
  \end{figure}

  The problem of foreground subtraction is challenging and previously used methods have been proven largely incomplete. 
One common way of treating the contamination problem has been to extrapolate the signal from frequencies where dust emission dominates ($\sim$ 350 GHz) to frequencies where CMB emission is important ($\sim$ 60-150 GHz). 
However, this extrapolation can become problematic. For example, in the presence of two clouds along the same line-of-sight,
the polarization at one frequency could be decorrelated compared to that at another frequency \citep{tassis_pavlidou}. The conditions for this decorrelation to be significant are:
1) the magnetic field of one cloud is significantly misaligned with that of the other and 2) the temperatures of the clouds are not identical.
  
Knowledge of the orientation of the magnetic field on the plane of the sky as a function of distance from the observer is necessary to address this effect. 
Thermal emission cannot provide this information because its intensity is the result of integration along the line-of-sight to infinity, and therefore distance information is lost. 
Stellar polarization, on the other hand, only traces the dust column out to the distance of the star. With enough measurements of stars of known distances tracing a similar sightline, one can reconstruct the plane-of-sky magnetic field orientation as a function of distance. 
Existing stellar polarization measurements are very sparse at high-latitudes, which are the regions targeted by CMB experiments. The optical polarization survey PASIPHAE\footnote{Polar-Areas Stellar Imaging in Polarization High Accuracy Experiment, http://pasiphae.science/} is being designed to address precisely this issue.
 
  In this work, we conduct a path-finding mini-survey for \pasiphae{} in three regions with very low dust content. We wish to determine the level of polarization that can be measured in regions with very low dust emission using a flux-limited sample of stars located within a very small area ($\sim 0.05$ square degrees). In contrast to previous works at high-latitude, this approach allows for determination of the average polarization properties of the ISM locally.
  
The surveyed regions and the observing strategy are presented in Sect. \ref{sec:observations}. A
 description of the reduction and instrument calibration are given in Sect. \ref{sec:reduction}, supplemented by appendix \ref{sec:appendix}. Our results are presented in Sect. 4. Our findings are discussed in Sect. \ref{sec:discussion} and our conclusions in Sect. \ref{sec:conclusions}.

\section{Observations}
  \label{sec:observations}
  
  Observations were conducted with the 1.3-m telescope at 
  Skinakas Observatory in Crete, Greece\footnote{http://skinakas.physics.uoc.gr/} using the RoboPol polarimeter \citep{king}. RoboPol is a four-channel imaging polarimeter, designed to simultaneously measure the relative Stokes parameters 
$q$ and $u.$  Each source in the 13$\arcmin \times$13$\arcmin$ field of view (FOV) is projected on four locations on the CCD. The central part of the FOV (2$\arcmin \times$2$\arcmin$) is shadowed by a focal plane mask whose purpose is to lower the background for the central target. For this project, we measure sources only in this masked region, to maximize measurement precision. We conducted all observations in the R-band.

 \begin{figure*}
    \centering
    \includegraphics[width=\textwidth]{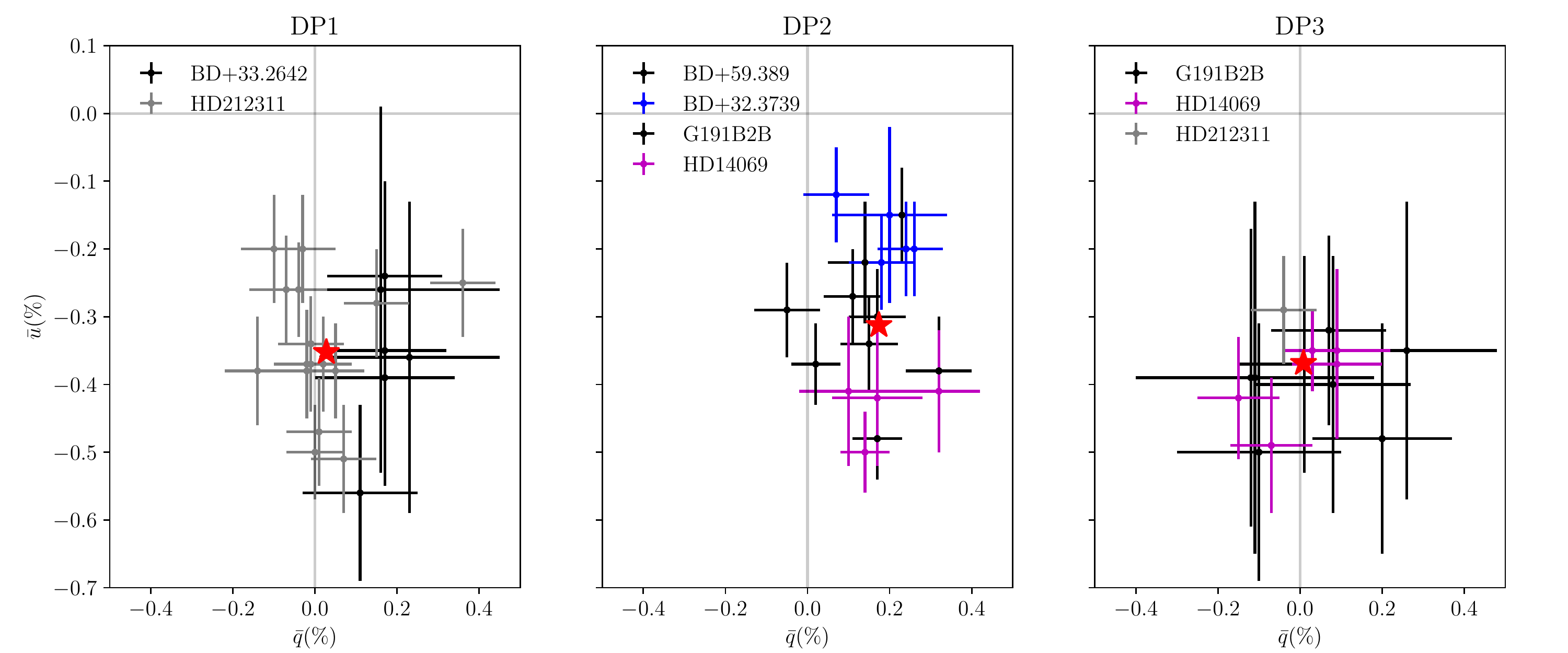}
    \caption{The $\bar{q}-\bar{u}$ plane of standard polarization stars. The red star 
             illustrated in each case is the weighted mean of measurements and 
             indicates the instrumental zero-point. The labels correspond
             to the names of the standard stars.}
         \label{fig:centroid_standards}
 \end{figure*}


  As stellar polarization depends on reddening, we find the mean E(B-V) towards the observed regions. We refer to the target regions as \textit{Dark Patches} (DPs). To this end, we use the \citet{lenz} reddening (hereafter LHD) map (Fig. \ref{fig:lenz_map}). This map was derived from HI emission using the HI4PI all-sky survey \citep{hi4pi} and covers $39\%$ of the sky with HI column densities $\rm N_{HI} < 4 \times 10^{20} cm^{-2}$. The map is provided in HEALPIX format with NSIDE = 1024 (pixel spacing $3.3\arcmin$). Figure \ref{fig:dark_patch_distrib} shows the distribution of E(B-V) in the LHD map. The resolution of this map ($16\arcmin$) is comparable to the size of our regions. Therefore, we assign a single value of E(B-V) to each DP: that of the E(B-V) map at the center of the region. The reddening of the target regions is shown with vertical lines at: 0.0063 mag (solid line, DP1), 0.0072 mag (dotted line, DP2) and 0.0118 mag (dash-dotted line, DP3).
We compare these values with an independent estimate of E(B-V) from the map of \cite{planck_ebv}. We assume $R_{v}=3.1$ \citep{schultz_wiemer} to convert the $A_{v}$ to E(B-V). Remarkably, (considering all the factors that contribute to the uncertainty) we find very small differences between these values and those of LHD. According to Planck, E(B-V) is 0.0083 mag in DP1,   0.0077 mag in DP2, and  0.0171 mag in DP3. In the E(B-V) map of Fig. \ref{fig:lenz_map}, 24, 94 and 1751 square degrees in total correspond to regions with E(B-V) lower than that of DP1, DP2, and DP3, respectively.

  
The centers in Galactic coordinates (l, b) and angular sizes of the three regions are: DP1; (124.7, 60.0), 15$\arcmin \times 15\arcmin$ ; DP2; (159.4, 49.0), $16\arcmin \times 16\arcmin$; and DP3; (191.1, 48.6), 13$\arcmin \times 13\arcmin$. The locations of the three surveyed regions are marked with red star symbols on the LHD map in Fig. \ref{fig:lenz_map}. 
   
Within each target region, we constructed flux-limited samples. Stars with 10.5 $\rm mag <R<16.5$ mag were selected from the USNO-B catalog \citep{usnob}. We discarded stars that would suffer from confusion with nearby sources. Due to observing-time limitations, our final samples are complete to 16.47 mag for DP1, 15.7 mag for DP2, and 16.25 mag for DP3 (with an additional star of 16.6 mag). 
    
  Observations took place in the period May-August 2015 for DP1, September-November 2015 for DP3, and October-November 2017 for DP2. The total observing time for the three regions was 27.5 hr for DP1, 16 hr for DP2, and 20 hr for DP3. In total we observed 68 stars from which 24 were in DP1, 23 in DP2, and 21 in DP3. For each star the exposure time was selected so that the statistical (photon-noise) error of $p$ would be comparable to the instrumental systematic error of 0.1\% (see Sect. \ref{sec:calibration}). Total exposure times for targets ranged from 2 to 165 mins, with $\sim$80\% of targets having an exposure time of $<50$ mins. 
  Polarization standard stars were observed each night for calibration. During the 2015 observing run, two standards were observed each night on most nights, and one standard on some nights, due to visibility/time constraints. During the 2017 observing run, from two to six standards were observed during each night, with the exception of one night when only one was observed.

%
%
\begin{table}
    \centering          
    \caption{Literature polarization of standard stars used for instrument
             calibration.}             
    \label{table:standards_table}      
    \begin{tabular}{c c c c l l l }   
      \hline\hline                     
       Name& $p$(\%) & $\chi(\circ)$ & Band & Ref \\ 
       \hline  
       BD+32.3739       & 0.025$\pm$0.017       & 35.79$^{\circ}$               & V &     1       \\
       G191B2B  & 0.061 $\pm$ 0.038     & 147.65$^{\circ}$      & V &   1\\
       HD212311         & 0.034 $\pm$ 0.021     & 50.99$^{\circ}$       &  V & 1\\
       HD14069  & 0.022 $\pm$ 0.019     & 156.57$^{\circ}$              &  V & 1\\
       BD+59.389        & 6.430$\pm$0.022       & 98.14$^{\circ}$ $\pm$ 0.10$^{\circ}$& R &     1       \\ 
       BD+33.2642       & 0.231$\pm$0.031       & 12.67$^{\circ}$               & V &     1       \\
       BD+33.2642   & 0.20$\pm$0.15  & 78$^{\circ}\pm$ 20$^{\circ}$  &R & 2\\            
       \hline          
    \end{tabular} 
    \tablebib{(1)\cite{schmidt}; (2) this work}
\end{table} 

\section{Data reduction}
  \label{sec:reduction}

  The data were reduced using the RoboPol pipeline \citep{king}, which performs aperture photometry of each source to measure the relative Stokes parameters $q$ and $u$ and their (statistical) uncertainties $\sigma_q$ and  $\sigma_u$, respectively. These are used to calculate the fractional linear polarization, $p$, and the electric vector 
  position angle (EVPA or $\chi$) through:
  \begin{equation}
    p = \sqrt{q^{2} + u^{2}},\hspace{2mm} \sigma_{p} = \sqrt{\frac{q^{2}
        \sigma^{2}_{q}+u^{2}\sigma^{2}_{u}}{q^{2}+u^{2}}},
    \label{eq:polarization_stokes}
  \end{equation}    
  \begin{equation}
    \chi=\frac{1}{2}\arctan\left( \frac{u}{q} \right), \hspace{2mm} \sigma_{\chi}=
         \frac{1}{2}\sqrt{\frac{u^{2}\sigma^{2}_{q}+q^{2}\sigma^{2}_{u}}
                                                            {(q^{2}+u^{2})^{2}}}.
    \label{eq:angles_stokes}
   \end{equation}
  We use the latest version of the pipeline which selects the aperture size according to the optimization method presented in \citet{panop2015}. 
  
Two modifications have been made after the publication of that paper. First, in \citet{panop2015}, the optimization was run separately on each of the four images of a target. However, this could potentially introduce artificial differences between the photometry of the stellar images, leading to low levels of spurious polarization (which could be significant for this work). To alleviate this, we use an aperture that is common for the pair of ordinary and extraordinary stellar images. In practice, we run the optimization on each stellar image separately, and then use an aperture size that is the mean of the two solutions for the pair of images used to calculate one Stokes parameter and likewise for the remaining pair of images.
Second, we recalibrate the aperture optimization for sources in the RoboPol mask. The optimal aperture for a stellar image is found by constructing its growth curve (intensity as a function of aperture size, $x$), fitting a polynomial $f(x)$ and solving the equation: $df/dx=\lambda f(x)$. The parameter $\lambda$ is calibrated using standard stars. A value of 0.02 was found for sources in the RoboPol field in \citet{panop2015}. For this work, we repeated the calibration of $\lambda$ for sources in the RoboPol mask and
found a value of 0.01. We use the same set of processing parameters for both polarization standard stars and DP target stars. 
  
    \label{sec:calibration}
    \begin{table}
      \caption{Instrumental zero-point for each observing run.}
      \label{table:instrumental_error}
      \centering                          
      \begin{tabular}{c c c }        
        \hline\hline                 
        Observing run & $ q_{inst}(\%)$ & $u_{inst}(\%)$\\    
        \hline                        
        DP1  & 0.03$\pm$ 0.12 & -0.35 $\pm$ 0.10 \\      
        DP2  & 0.17$\pm$ 0.09 & -0.31$\pm$ 0.12\\
        DP3  & 0.01$\pm$ 0.12 & -0.37$\pm$ 0.06\\
        \hline                                   
      \end{tabular}
    \end{table}

The instrument calibration involves using polarization standard stars to (a) determine the polarization zero-point, (b) estimate the uncertainty of this determination (systematic error), and (c) identify the rotation of the coordinate system compared to the standard (IAU) reference frame. In the majority of works using RoboPol data, calibration is done using an instrument model, which is constructed by scanning standard stars across the instrument FOV \citep{king}. In this way, the polarization zero point is found for every point on the CCD. This approach, however, only provides an estimate of the systematic uncertainty for sources observed in the field of the instrument \citep{panop2015}. Our sources were observed in the mask, so as to minimize systematic uncertainties. For this reason, and to eliminate unknown uncertainties in the model determination, we do not make use of an instrument model for the calibration of the DP sources. Instead, we use the measurements of the standards to find the weighted mean instrumental zero-point and its corresponding uncertainty. 

        \begin{figure*}
      \centering
      \includegraphics[width=\textwidth]{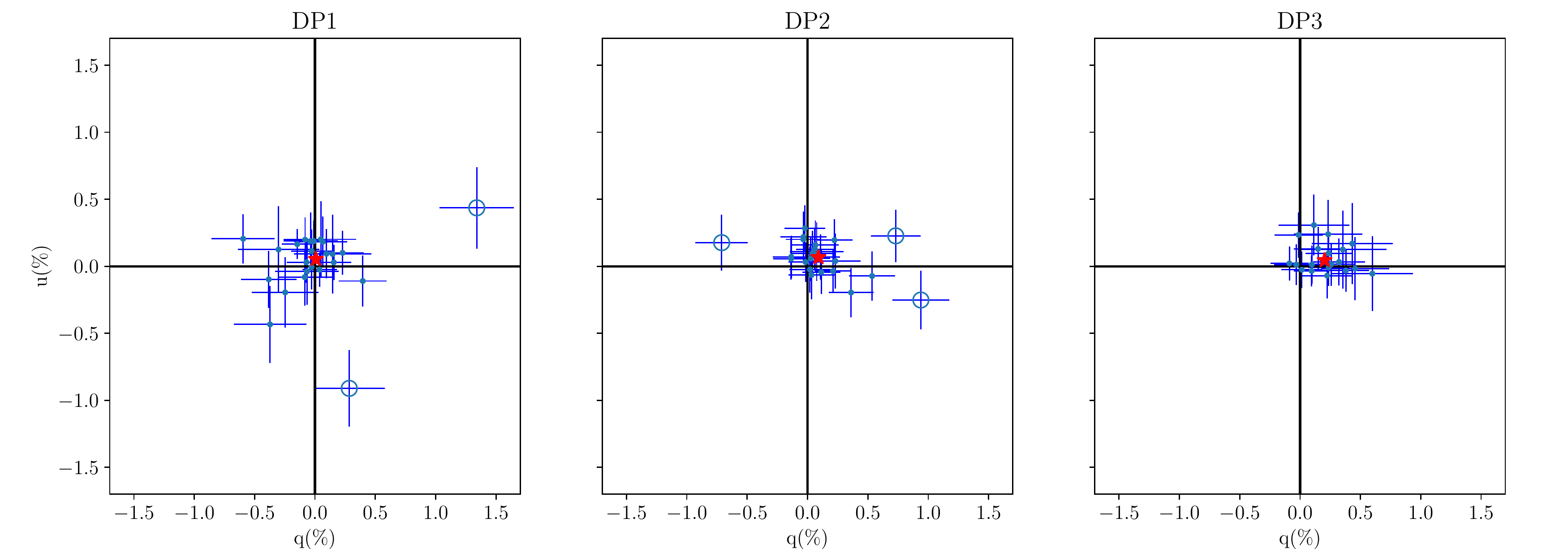}
      \caption{Measurements of Stokes parameters,
$q$ and $u,$ of stars in the DP fields, after instrumental calibration (Sect. \ref{sec:calibration}) shown with blue points. Error bars correspond to the combined statistical and systematic uncertainties. The red star in each panel corresponds to the weighted mean of $q$ and $u$ measurements in each DP ($q_{mean}$, $u_{mean}$ in Table \ref{table:stokes_weighted}). Empty circles denote the outliers defined in Sect. \ref{subsec:outliers}.}
          \label{quplane}
    \end{figure*}

        The set of polarization standards used for calibration are shown in Table \ref{table:standards_table}, along with their literature values. Though our observations were conducted in the R$-$band, not all standard stars have reference values in this band. However, most stars which do not have a measurement in the R$-$band are polarized at a level well below the typical systematic error of our instrument (0.1\% in the mask). Assuming that their polarization is interstellar, differences between the V$-$ and R$-$ bands will be negligible for our purposes. The only exception is the star BD+33.2642, which has a reference value of $p=$ 0.2\% in V. We present a determination of the R$-$band value of this standard in Appendix \ref{sec:appendix}. We use this value for the following calibration steps.
    
    To find the instrument zero-point, we first subtract the literature value ($q_{lit}$) from each measurement of a standard star ($q_{obs}$): $\bar{q}=q_{obs}-q_{lit}$ and $\bar{u}=u_{obs}-u_{lit}$. The zero-point of the instrument $q_{inst}$ ($u_{inst}$) is the weighted mean of all $\bar{q}$ ($\bar{u}$). 
    
    The uncertainty of the zero-point reflects the level of systematic error. For its determination, it is of critical importance to take into account all the factors that can contribute to this uncertainty. These include (a) possible intrinsic variability of the standard polarization stars, (b) errors in the determination of the literature value of a certain star (if one makes use of multiple stars for the determination), (c) (spatio-temporal) variability of the sky conditions, and (d) (spatio-temporal) variability of the instrument behavior.

    Although the zero-point is found using the weighted mean of measurements, the standard error on the mean cannot capture all the aforementioned factors. In the limit of a very large number of measurements of standard stars, the standard error would tend to zero, even though these sources of error would still be at play.  
    In order to properly quantify the aforementioned effects, we assign the uncertainty on the zero-point to be the standard deviation of the $\bar{q}$ measurements (and correspondingly for $\bar{u}$). This is a conservative approach compared to the standard error on the mean and is more likely to err on the side of caution, that is, it is likely to overestimate the systematic uncertainty.
    
    Figure \ref{fig:centroid_standards} shows the $\bar{q}-\bar{u}$ plane of the observed standards. Measurements of standards during the 
    observation time span of DP1 are illustrated in the left panel, in the middle panel for DP2 and in the right for DP3. It is clear that for all observing epochs, the instrument biases the observations towards more positive $q$ and more negative $u$ values. The instrument zero-point is marked with a red star. In Table \ref{table:instrumental_error}, we present the instrument zero-point ($q_{inst}$, $u_{inst}$) for all the regions, and its corresponding uncertainty. We find the systematic uncertainty to be at the level of 0.1\%.
    
In addition to the zero-point shift, instrumental effects may also result in a rotation of the $q - u$ plane compared to the standard (EVPA zero at north, increasing towards east). 
In practice, to calculate the instrumental rotation we select polarized standards and correct their values for the zero-point shift found previously. Then, we find the average EVPA from these corrected $q,u$ ($\chi_{pol,mean}$) and subtract from it the literature value of the EVPA: $\chi_{pol,mean} - \chi_{lit}$. 

This rotation is very small for RoboPol, with measurements of polarized standards in 2017 and 2015 placing it at $0.5^\circ \pm 0.1 ^\circ$, much smaller than statistical uncertainties for all values of EVPA quoted in this work. 

\section{Results}

\label{sec:results}

        \begin{figure*}
      \centering
      \includegraphics[width=\textwidth]{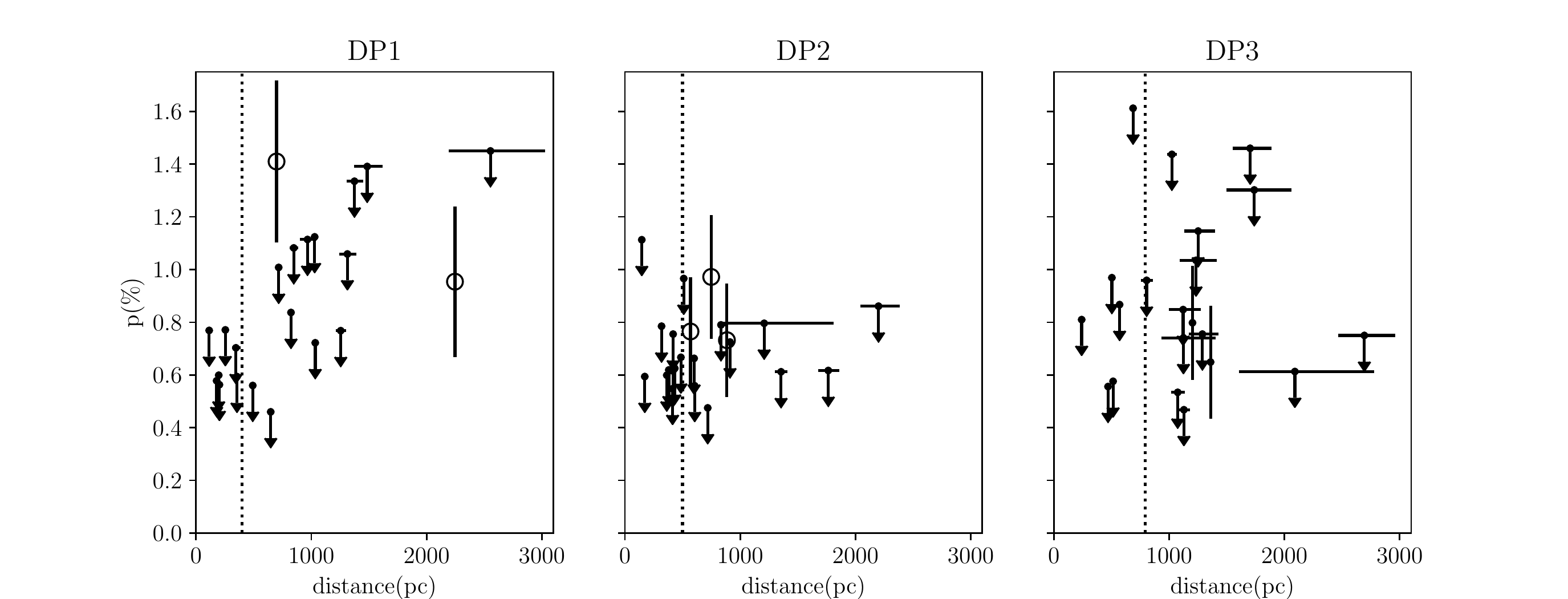}
      \caption{Degree of polarization vs. distance for all the stars with measured parallaxes by Gaia in the DP fields. Downward-pointing arrows represent the $3\sigma$ upper limit in polarization fraction when $p/\sigma_{p}<3$. Some distance error bars are too small to be seen. The dotted vertical lines correspond to the distance at which maximum E(B-V) is reached along the line-of-sight of each region according to \cite{green_map}. Empty circles denote the outliers defined in Sect. \ref{subsec:outliers}. }
      \label{fig:tomography}
    \end{figure*}
                
    \subsection{Measurements on the $q-u$ plane}

    After reducing the data with the RoboPol pipeline, we correct each measurement of a DP target for the instrumental zero-point, and propagate the statistical and systematic uncertainty to the final result. We plot the corrected measurements on the $q-u$ plane in Fig. \ref{quplane}. As is clear from Eqs.~\ref{eq:polarization_stokes} and \ref{eq:angles_stokes}, $p$ measures the offset from the origin, while $\chi$ measures the (half) angle with respect to the line of positive $q$, in the counter-clockwise direction. 
    
    In each DP, most of the measurements cluster near the origin. Only two stars in DP1, three  in DP2, and two in DP3 have signal-to-noise ratio (S/Ns) in $p$ higher than $3$. We want to investigate if this is due to the fact that the observed stars are nearby, therefore meaning the polarization does not trace the full extinction across the line-of-sight. We use \textit{Gaia Data Release 2} \citep{gaia} to derive the distances of the observed stars. In DP1, there are parallaxes for 22 stars out of 24, in DP2 for 22 out of 23, and in DP3 parallaxes exist for all the stars we observed. However, inverting parallaxes is not a reliable method for inferring distances \citep{luri_brown}. For this reason we use the distances published by \cite{bailer_jones}\footnote{http://gaia.ari.uni-heidelberg.de/tap.html}.
    
        In Fig. \ref{fig:tomography}, we present the degree of polarization ($p$) versus distance for all the stars with associated distances. The majority of measurements are non-detections ($S/N$ in $p$ less than 3) and we present their upper limits (black arrows) within $3 \sigma$. For stars with $S/N>3,$ we present their observed uncertainties. For several stars, the distance uncertainty is negligible and is not visible. We compare this with the line-of-sight distribution of E(B-V) for each DP using the latest version of the three-dimensional (3D) Galactic E(B-V) map\footnote{http://argonaut.skymaps.info/} of \cite{green_map}. This map's beam size ranges from $3.4\arcmin$ (for high extinction regions) to $13.7\arcmin$ (for low extinction regions). We query the map for the E(B-V) as a function of distance across the line-of-sight at which each DP is centered. In DP1, DP2, and DP3 the plateau of maximum E(B-V) is reached at distances of $398$, $501$, and $794$ pc, respectively (shown with vertical dotted lines in Fig. \ref{fig:tomography}). Altogether, in all DPs, the majority of stars are far enough to trace the full extinction across the line-of-sight. Therefore, our measurements are tracing the line-of-sight-averaged polarization.
        
    Our approach of observing multiple stars within a small area enables us to infer the mean interstellar polarization towards these regions, even though we do not have a significant detection of $p$ for individual stars. This information is encoded in the observed anisotropy towards a certain direction on the $q-u$ plane. Starting with DP3, the clustering of $q$, $u$ measurements towards the first quadrant indicates a non-zero mean polarization; the location of this clustering is related to the mean direction of the local polarization of this region. As we move from DP3 to DP1, the anisotropy of measurements around the origin becomes less pronounced. In DP1, most of the measurements are distributed roughly isotropically around zero, appearing consistent with a non-polarized region at the accuracy level of our instrument. DP2 measurements show some clustering towards the first and fourth quadrant, although the $q$ and $u$ measurements are not as anisotropic as in DP3. In Sect. \ref{sec:mean}, we calculate the mean $p$ and $\chi$ of each region using the weighted mean $q$ and $u$ (marked with a red star in Fig. \ref{quplane}).
     
    \subsection{Search for indicators of intrinsic polarization}  
    \label{subsec:outliers}
    While most stars are clustered on the $q-u$ plane, there are some prominent outliers; these are the measurements for which $p$ is further than $3\, \sigma_p$ from the mean $p$. More specifically, in DP1 there are two outliers in the first and fourth quadrants of the $q-u$ plane. In DP2 there are three outliers: one in the second quadrant, located opposite the majority of the measurements, one in the fourth quadrant, and the last one in the first quadrant. These measurements are denoted with empty circles in Figs. \ref{quplane} and \ref{fig:tomography}. Inspecting the latter, we cannot attribute the high $p$ we measured to the fact the stars are far away; except for the outlier in DP1 located at a distance of $2100$pc.
    
    One possible reason for the existence of these outliers could be that they are intrinsically polarized. If this is the case, their measurements should be excluded from our analysis of the  properties of the mean interstellar polarization in the surveyed regions. We therefore searched for complementary information on the sources that could help us judge whether they are potentially  intrinsically polarized. 
    
    One type of source that exhibits intrinsic polarization is an active galactic nucleus (AGN) \citep{angel}. AGNs are easily distinguished from stars by their non-black-body multiwavelength emission. In our search, we used VOSA\footnote{http://svo2.cab.inta-csic.es/theory/vosa/} \citep{bayo}, a tool which uses historical multi-band photometric data in order to construct the spectral energy distribution (SED) of a source. We fit the simplest stellar spectral model of VOSA to the outlier sources and find that all are consistent with a black body spectrum. We also computed their effective temperatures, as these can indicate if a star is young (and therefore likely to have a polarization-inducing circumstellar disk). However, all the temperatures found are typical of main sequence stars.
    
    Our second approach to search for intrinsic variability utilizes data from the second data release of the Catalina Sky Surveys\footnote{http://nesssi.cacr.caltech.edu/DataRelease/} \citep{drake}. For each of our targets, we inspected the seven-year photometric light curves and found no sign of intrinsic (photometric) variability. In addition, we checked the observed B-V colors of the stars in our sample to see if there could be any Be star candidates. We used the SDSS G-R color and converted to B-V 
according to \cite{sebastian}. We found no stars with negative B-V values (that would be consistent with O or B types). Finally, according to the Besancon stellar population synthesis model of the Galaxy \citep{besancon}, no stars of type O-A are found within our survey magnitude range in the observed regions.
    We therefore proceed using the entire sample of observed targets.
    
    \subsection{Mean polarization}
        \label{sec:mean}
    
   \begin{table*}
      \caption{Average interstellar polarization properties in each DP. Left two columns: weighted mean Stokes parameters ($q_{mean}$, $u_{mean}$). Middle two columns: fractional linear polarization, $p_{mean}$ (calculated from $q_{mean}$, $u_{mean}$ according to Eq. \ref{eq:polarization_stokes}) and its debiased estimate $\hat{p}_{mean}$. Right column: polarization angle $\chi_{mean}$ derived from $q_{mean}$, $u_{mean}$, using Eq. \ref{eq:angles_stokes}.}
      \label{table:stokes_weighted}
      \centering                          
      \begin{tabular}{c c c c c c}        
        \hline\hline                 
        &$q_{mean} \, (\%)$  &  $u_{mean} \, (\%)$  & $p_{mean} \, (\%)$ & $\hat{p}_{mean} \, (\%)$& $\chi_{mean} \,(^{o}) $\\    
        \hline                        
        DP1 & 0.007 $\pm$ 0.041  & 0.053 $\pm$ 0.037   &  0.054$\pm$ 0.038 & $0.0 $& 41 $\pm$ 22\\      
        DP2 & 0.091 $\pm$ 0.036  & 0.066 $\pm$ 0.036   & 0.113$\pm$ 0.036 & 0.107 & 18$\pm$ 9\\
        DP3 &0.203 $\pm$ 0.045  &  0.045 $\pm$ 0.037     & 0.208$\pm$ 0.044 & 0.203 & 6$\pm$ 5\\
        \hline                                   
      \end{tabular}
     \end{table*}

    In order to measure the mean fractional linear polarization, $p_{mean}$, and the mean EVPA, $\chi_{mean}$, for each DP, we computed the weighted mean of the $q$ and $u$ measurements of each region. We obtain two single values ($q_{mean}$ and $u_{mean}$) for each region, and we apply Eqs. \ref{eq:polarization_stokes} and \ref{eq:angles_stokes} to derive $p_{mean}$ and $\chi_{mean}$. We follow this approach for two reasons. First, it minimizes the contribution to the mean of the aforementioned outliers which might not be consistent with ISM polarization. Second, this approach avoids the bias of individual measurements. The majority of our measurements have low $p/\sigma_p$ and are therefore biased towards higher values of $p$ \citep[e.g.,][]{simmons_stewart,vaillancourt}. Therefore, if we were to compute the $p_{mean}$ of each region by averaging individual stellar $p$, the $p_{mean}$ value would be overestimated. 
      
    In Table \ref{table:stokes_weighted}, we present the mean values of the Stokes parameters and $p$ and $\chi$ in each region. The $p_{mean}$ in DP1 lies within $1.4\sigma$ of the origin, while that of DP2 is $3.1\sigma$ from zero, and that of DP3 is $4.7\sigma$ from zero. Therefore, we have measured significant polarization in DP3 and DP2 but not in DP1. We consider the $p_{mean}$ of DP2 to be a lower bound on the  detectable mean polarization that can be achieved using our methodology.
    
    The isotropy of $q$, $u$ measurements of DP1 about the origin is consistent with the non-detection of polarization. As we move from DP2 to DP3, the anisotropy of $q$, $u$ measurements becomes more prominent. We elaborate more on the anisotropy of the measurements with regard to the significance of the $p_{mean}$ in the DPs in Sect. \ref{sec:symmetry}.
    
    \subsection{Intrinsic spread in the distribution of Stokes parameters}
    
    Due to the distribution of stars in three dimensions, different stars could be tracing different materials along the line-of-sight. As a result, deviations from the mean polarization (or equivalently from $q_{mean}$ and $u_{mean}$) can arise. This means that there is an underlying intrinsic distribution of $q$ and $u,$ respectively, which needs to be fully characterized. Except from $q_{mean}$ and $u_{mean}$, our strategy of measuring a large number of stars within a small region of the sky allows us to constrain the intrinsic spread of the distributions of $q$ and $u$. We proceed to obtain such constraints only for the two regions where we have detected significant $p_{mean}$ (DP2, DP3).
        
     The spreads (sample standard deviations) of the observed $q$ and $u$ are $s_{q, obs}=0.31 \%$ and $s_{u, obs}=0.13 \%$ for DP2, and $s_{q, obs}=0.22 \%$ and $s_{u, obs}=0.12 \%$ for DP3. The errors in each individual stellar measurement are comparable to the observed standard deviations of the distributions. This means that the spreads of these distributions are determined by the uncertainty of our measurements and not by a physical process. 

        \begin{figure}
      \centering
      \includegraphics[width=\hsize]{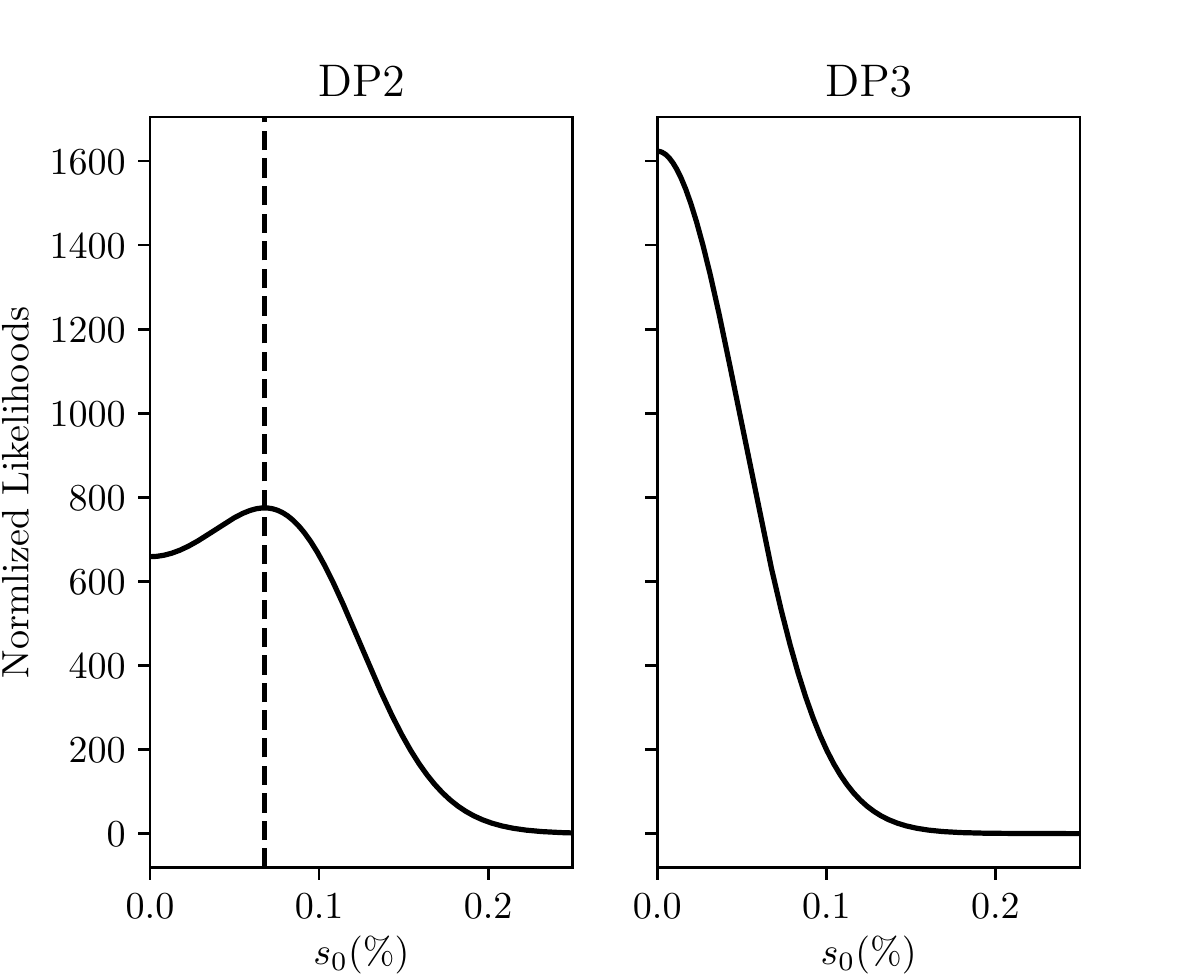}
      \caption{Normalized likelihood $L$ as a function of the intrinsic spread $s_0$ of the distributions of Stokes parameters for DP2 (left) and DP3 (right). A vertical line shows the maximum likelihood $s_0$. The $99\%$ confidence intervals for DP2 and DP3 are $s_{0} \leq 0.187\%$ and $s_{0} \leq 0.127\%$, respectively.}
      \label{fig:likelihood}
    \end{figure}
    
    \begin{figure}
      \centering
      \includegraphics[width=\hsize]{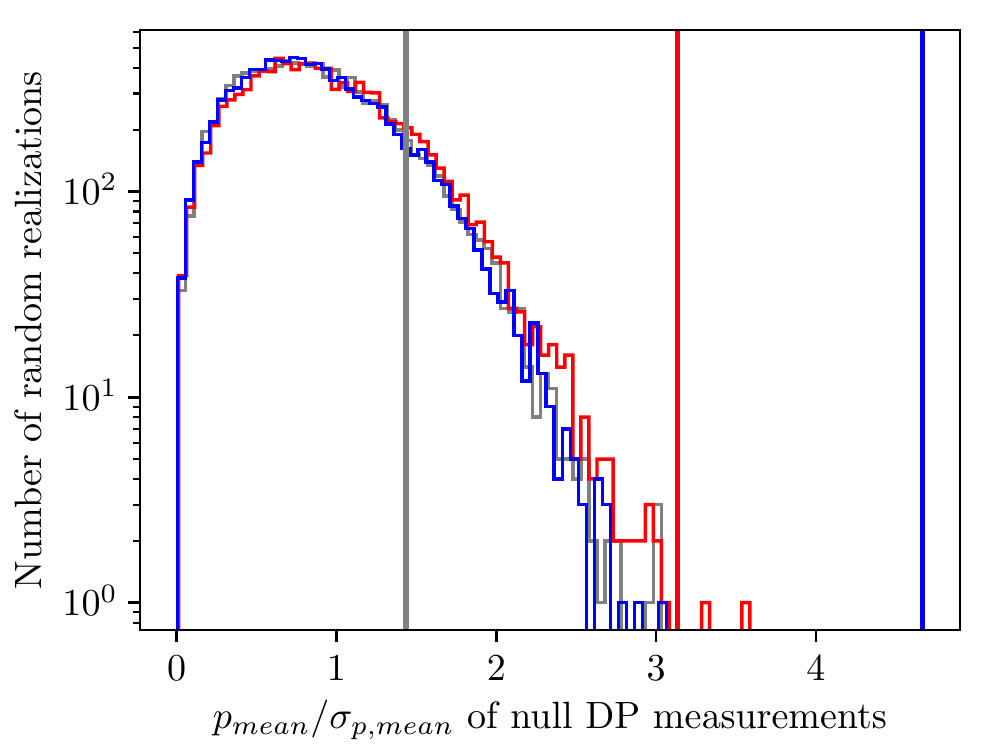}
      \caption{Distribution of the ratio of weighted mean $p$ over its error for random realizations of hypothetically unpolarized DP stars. The vertical lines show the actual ratio for our DP measurements. Colors: gray - DP1, red - DP2, blue - DP3.}
      \label{fig:nulltest}
    \end{figure}
    
    We use
a Bayesian approach to estimate the intrinsic spread of the  distributions. First we make the assumption that the intrinsic $q, u$ of the stars in a region are normally distributed about $q_{mean}, u_{mean}$ with equal spreads $s_{q,0}=s_{u,0}=s_{0}$. We also use the fact that measurements of the Stokes parameters of an individual star have Gaussian uncertainties. The likelihood of observing $N$ stars with $q_{obs,i}, u_{obs,i}$ ($i = 1,...N$) and measurement uncertainties $\sigma_{q, obs, i}$ and $\sigma_{u, obs, i}$ if the intrinsic spread is $s_0$ is:
    
    
    
    \begin{multline}
      L=\bigg( \prod_{i=1}^{N} \frac{1}{2 \pi \sqrt{(s_{0}^{2}+\sigma_{q,obs,i}^{2}) (s_{0}^{2}+\sigma_{u,obs,i}}^{2}) } \bigg) \times \\
        exp \bigg[ -\frac{1}{2} \bigg( \sum_{i=1}^{N}\frac{(q_{obs,i}-q_{mean})^{2}}{\sigma_{q,obs,i}^{2}+s_{0}^{2}} + 
                                                      \frac{(u_{obs,i}-u_{mean})^{2}}{\sigma_{u,obs,i}^{2}+s_{0}^{2}}\bigg) \bigg].
      \label{likelihood_total}
    \end{multline}
    The analytical proof of the likelihood function can be found in the appendix of \cite{venters}. 
    
    To obtain an estimate of the true $s_0$ we search for the value of $s_0$ that maximizes the likelihood function in Eq. \ref{likelihood_total}. Figure \ref{fig:likelihood} shows the likelihood $L$ as a function of $s_{0}$ for DP2 (left) and DP3 (right).
The functions are normalized so that the area under each curve is equal to 1. A vertical dotted line shows the value of $s_{0}$ which maximizes the likelihood. In the case of DP2, this corresponds to $s_{0}=0.07\%$, while for DP3 the likelihood peaks at $s_{0}=0\%$. Both likelihood functions in Fig. \ref{fig:likelihood} are bounded by zero and it is therefore not possible to obtain symmetric bounded confidence intervals on $s_0$. We can, however, place upper limits on $s_0$ for both regions: the $99\%$ upper confidence interval is $s_{0} \leq 0.187\%$ for DP2 and $s_{0} \leq 0.127\%$ for DP3.

For the purpose of this work, deriving upper limits on $s_0$ is sufficient. In DP3, we find  the maximum likelihood $s_0$ to be zero. This is reasonable since the majority of measurements are within $\sim 1\sigma$ of the $q_{mean}$ and $u_{mean}$. The situation in DP2 is altered mainly because of the presence of the outliers.


        \subsection{Confidence of the measured mean fractional linear polarizations}
    \label{sec:stattests}
    Due to the low signal in the DPs, we must assess the confidence that we can place on the presented mean
    interstellar polarization towards the three regions. We do this by quantifying the likelihood of a false detection,
    that is, the likelihood that the signal could result simply from uncertainties in our analysis. We use two observables to test this null hypothesis: the anisotropy of 
    measurements on the $q - u$ plane (\ref{sec:symmetry}) and the ratio of the mean fractional linear polarization over its uncertainty, $p_{mean}/\sigma_{p,mean}$ (\ref{sec:null}).
    
    \begin{figure}
      \centering
      \includegraphics[width=\hsize]{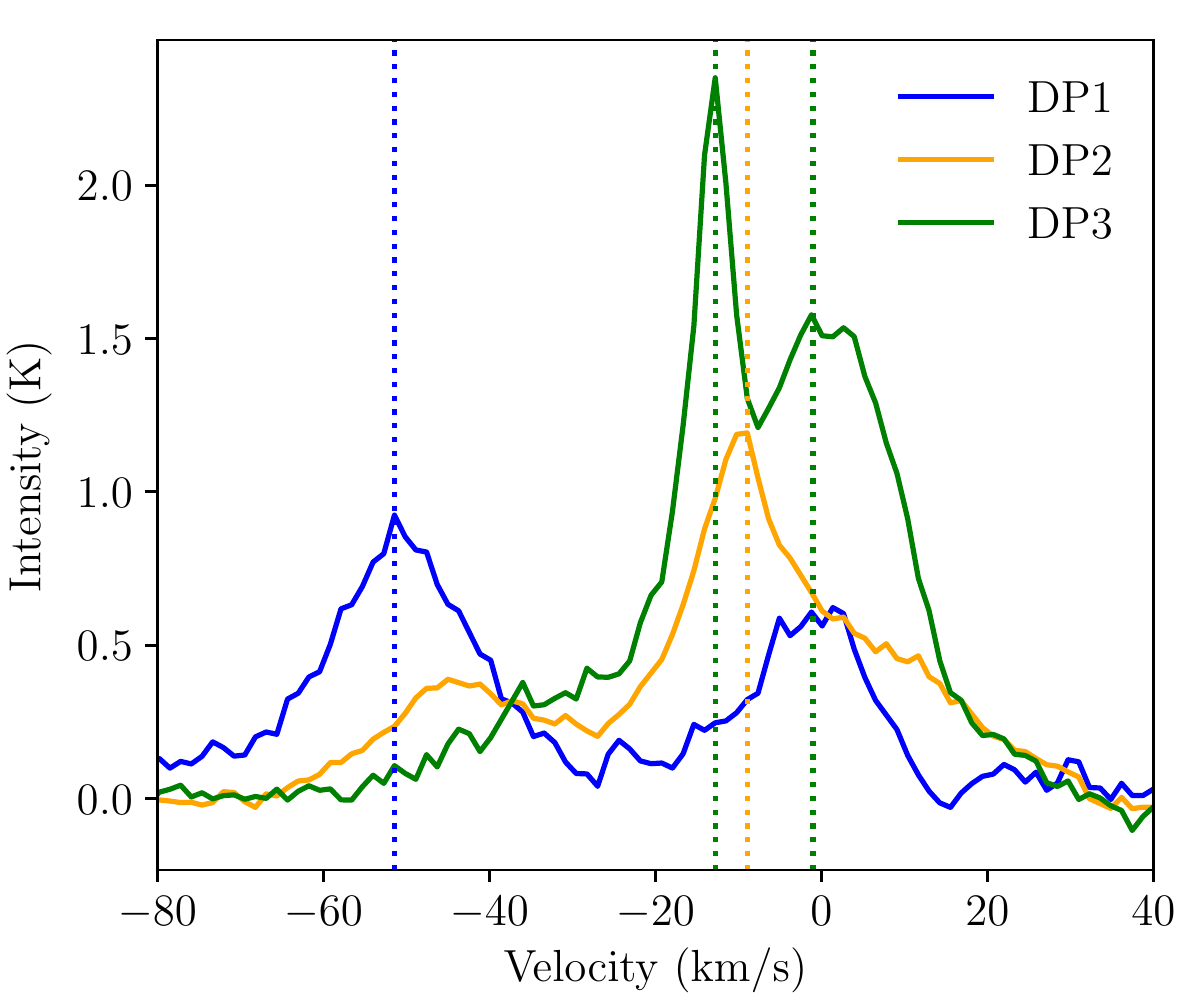}
      \caption{HI line intensity as a function of radial velocity for regions of  $10\arcmin$ in size, centered on the DPs. Vertical dotted lines correspond to the velocities at which the intensity peaks.}
      \label{fig:spectra}
    \end{figure}

    \subsubsection{Significance of $q-u$ plane anisotropy}
    \label{sec:symmetry}
    In DP1, there are seven measurements located in the first quadrant, eight in the 
    second, six in the third and three in the fourth. In DP2, there are nine measurements
    in the first, seven in the second, zero in the third and seven in the fourth quadrant.
    For DP3, the numbers are ten for the first, three for the second, zero for the third 
    and eight for the fourth quadrant. To quantify how anisotropically the 
    measurements are distributed around zero, we define an isotropy parameter $\kappa$ as the ratio of the number of $q, u$ measurements detected in the quadrant with the least measurements over the number of points detected in the quadrant with the most measurements. If a quadrant is empty, $\kappa=0$, while if all the measurements are isotropically 
    distributed, $\kappa=1$. For DP3 and DP2, $\kappa=0$ because the third 
    quadrant in each region has no measurements. In DP1, the fourth quadrant has 
    the least number of points, three, while the second has the most, eight, and therefore $\kappa=0.375$.
    
In order to investigate how probable it is to reproduce the observed anisotropies from unpolarized stars, we performed Monte Carlo simulations to obtain the probability distribution of $\kappa$ under the null hypothesis that all stars are unpolarized, ($q_{true}$, $u_{true}$) = (0,0).
    The $q_{i}$ and $u_{i}$ measurements follow Gaussian distributions centered on  $q_{{\rm true},i}$ and $u_{_{\rm true},i}$ with standard deviation $\sigma_{q,i}$ and $\sigma_{u,i}$ respectively; with i=1, $\dots$,N where N is the number of stars. Assuming all stars are unpolarized ($q_{{\rm true,}i}=0$ and $u_{{\rm true},i}=0$), we created mock observations  by drawing random values for $q$, $u$ from zero-centered Gaussians with standard deviations equal to the observational uncertainties of each star. We produced a sample of N mock ($q_{mock,i}$, $u_{mock,i}$) sets that match the number of stars measured in each region, and we computed the parameter $\kappa$. We repeated this process $10^6$ times, produced the distribution of mock $\kappa$, and  then compared it with the $\kappa$ value obtained from the real data. For DP1, we find that 40\% of the mock $\kappa$ are smaller than the observed one. For DP2 and DP3, this probability is 0.96\% and 0.95\%, respectively.
    
    We conclude that the anisotropy of observations on the $q,u$ plane in DP1 is consistent with the anisotropy produced by stars with zero ISM polarization (unpolarized). In the other DPs, the observed anisotropy could not be produced by unpolarized stars, with confidence more than 99$\%$.
 
 \subsubsection{Significance of $p_{mean}/\sigma_{p,mean}$}
 \label{sec:null}

In this section, we calculate the probability of measuring the observed weighted mean $p$ S/N ($p_{mean}/\sigma_{p,mean}$), if all stars in each DP field were unpolarized,  given the uncertainties in our analysis.  These uncertainties include the photon noise of individual DP target star measurements, but also uncertainties in the measurements of the standard stars that are used for our calibration (zero-point offset correction). We perform the calculation as follows. 

For each standard star used in the calibration of a DP, we draw a mock observation from a Gaussian centered on the existing measurement with a standard deviation equal to the statistical error of the measurement. We then compute the weighted mean of this mock set of standard observations (as in Sect.  \ref{sec:calibration}). This is the zero-point, around which any unpolarized star should lie. We now generate mock observations of the stars in the DP, assuming they are unpolarized. In practice, for each star we draw a value ($q$ and $u$) from a Gaussian centered on the zero-point that we had just calculated, with a standard deviation equal to the observed photon-noise error of the star. We now have mock observations of hypothetical zero-polarized stars in the DP. Next, we correct each mock star measurement for the instrumental polarization using the generated zero-offset and standard deviation, as in Sect. \ref{sec:calibration}. We then calculate the weighted mean and error of these corrected mock measurements. This process is repeated $10^4$ times. 
 
We construct the distribution of $p_{mean}/\sigma_{p,mean}$ from this test for each DP in Fig. \ref{fig:nulltest}. The distributions for all three DPs are very similar.  There is a peak at 0.8 and a long tail that extends out to 3.1 (3.6 for DP3). The observed $p_{mean}/\sigma_{p,mean}$ in the DPs (from Table \ref{table:stokes_weighted}) are shown with vertical lines. The $p_{mean}/\sigma_{p,mean}$ of DP1 falls well within the spread of the distribution, showing that it can be completely explained by the uncertainties present in our analysis. That of DP2 falls at 3.13, slightly higher than the maximum value of the $10^4$ DP2 zero-polarization realizations. The $p_{mean}$ S/N of DP3 is much larger than the corresponding maximum value of $10^4$ DP3 zero-polarization realizations. 

Both tests in this section are in agreement that DP2 and DP3 have yielded significant detections of the mean polarization, in contrast to DP1.

 \subsection{Characteristics of HI emission towards the DPs}
    \label{sec:hi}

    Having assessed the confidence of our measurements of the mean polarization in the three regions, we can now study the source of the signal; that is the diffuse atomic medium, as traced by HI line emission. 
    
    We use spectral data cubes from the first data release of the Effelsberg-Bonn HI survey \citep{winkel}, which have a beam size of 10$\arcmin$. For each DP, we locate the pixel in the HI data that corresponds to the center of the observed field. We then average the HI spectra of this pixel and its eight nearest neighbors (which yields an averaging area $\sim 10 \arcmin$ in width). Figure \ref{fig:spectra} shows the (averaged) HI spectra for all DPs (blue for DP1, orange for DP2, green for DP3). A vertical dotted line shows the velocity, $v_{max}$, at which each spectrum intensity is maximum. For DP1 and DP2, $v_{max}$ is -52 $\rm km/s$ and -9 $\rm km/s,$ respectively. For DP3, we show both the primary and the secondary peaks which correspond to velocities of -13 $\rm km/s$ and -1 $\rm km/s,$ respectively.
    
        \begin{figure}
   \centering
   \includegraphics[width=\hsize]{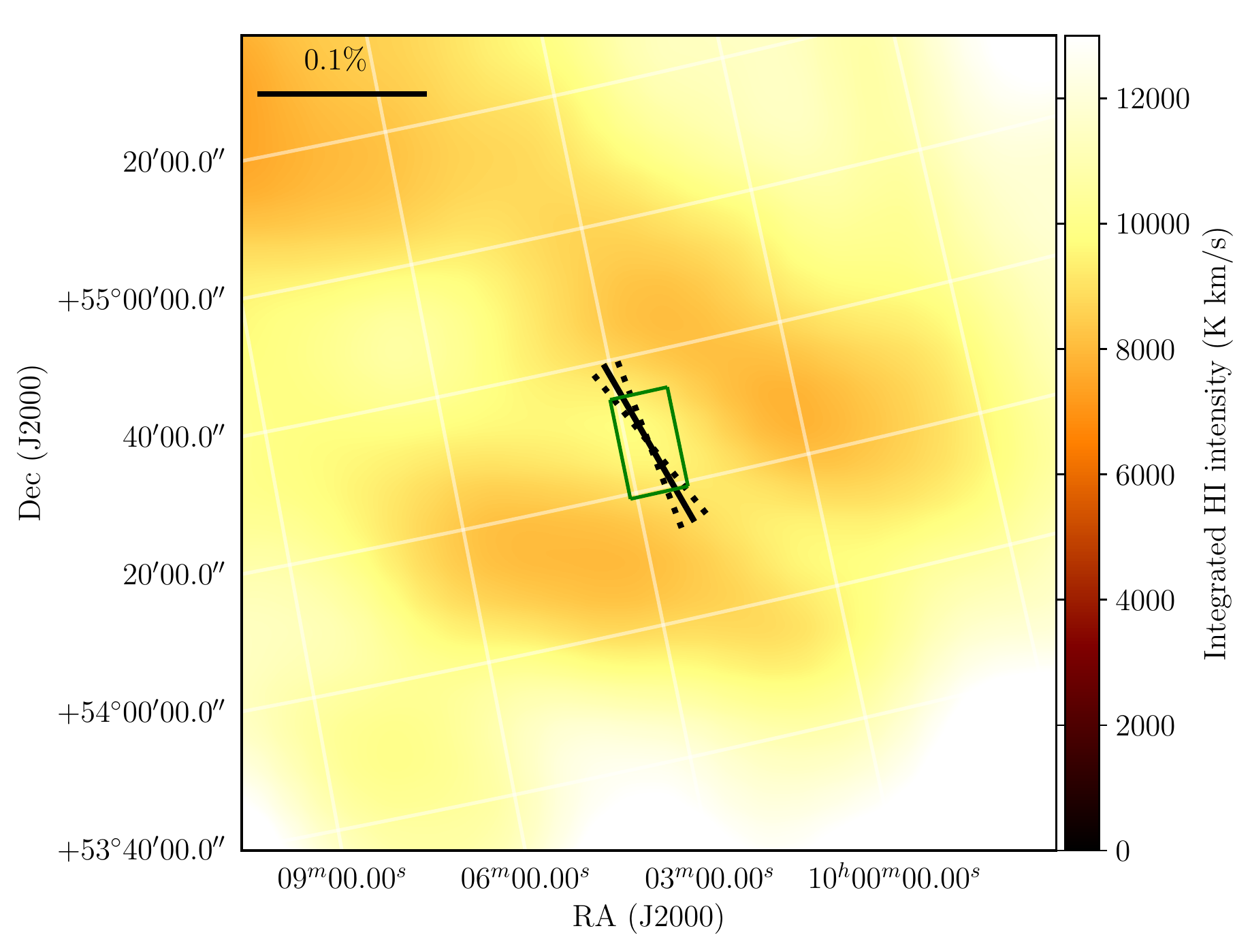}
      \caption{Mean polarization orientation over-plotted on an image of HI integrated intensity centered on DP2. Velocities are integrated around -9 km/s. The green box marks the surveyed region. The segment in the top-left corner is for scale in $p$.}
         \label{fig:hi_data_dp2}
   \end{figure}

   \begin{figure}
        \begin{minipage}[b]{\linewidth}
        \includegraphics[width=\linewidth]{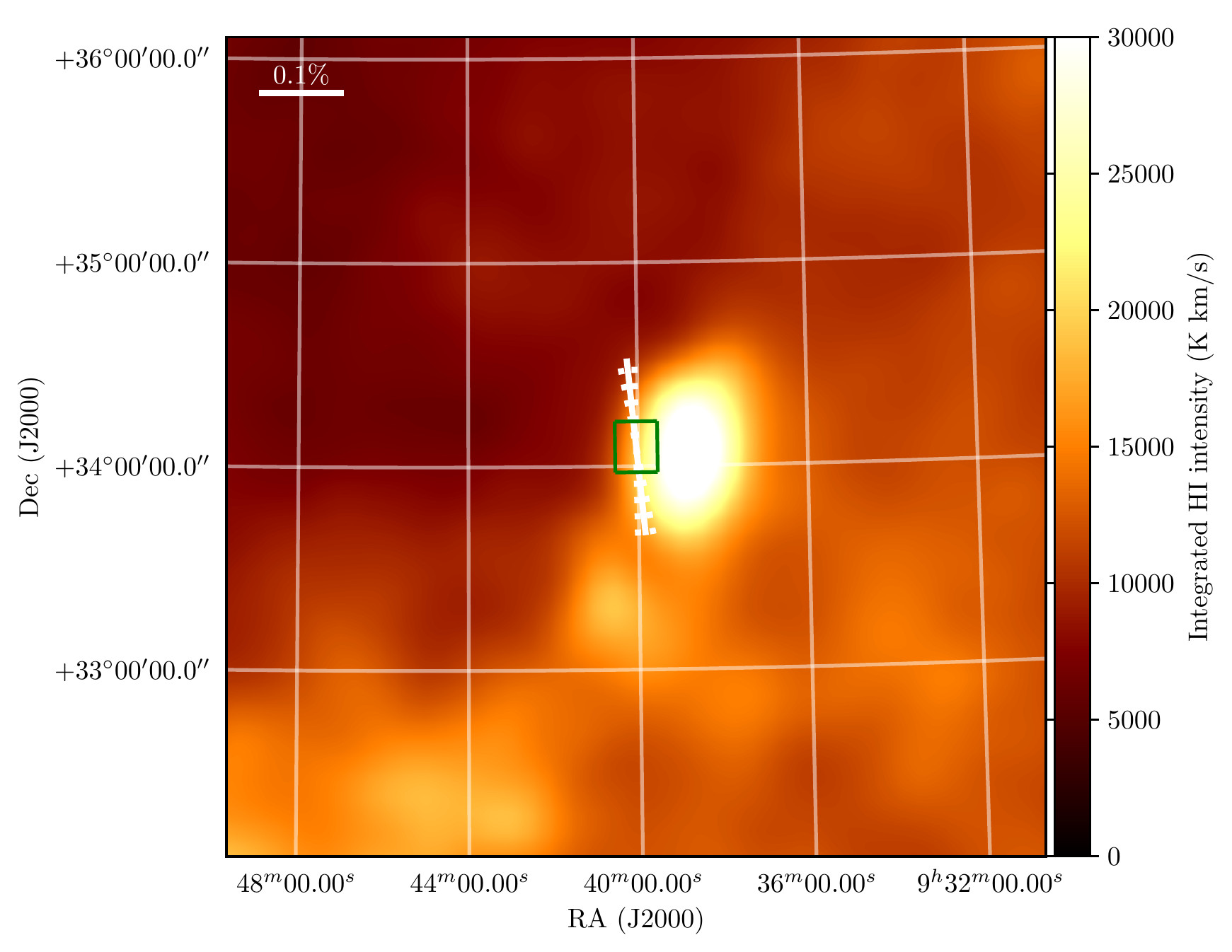}
    \end{minipage}
    \begin{minipage}[b]{\linewidth}
    \includegraphics[width=\linewidth]{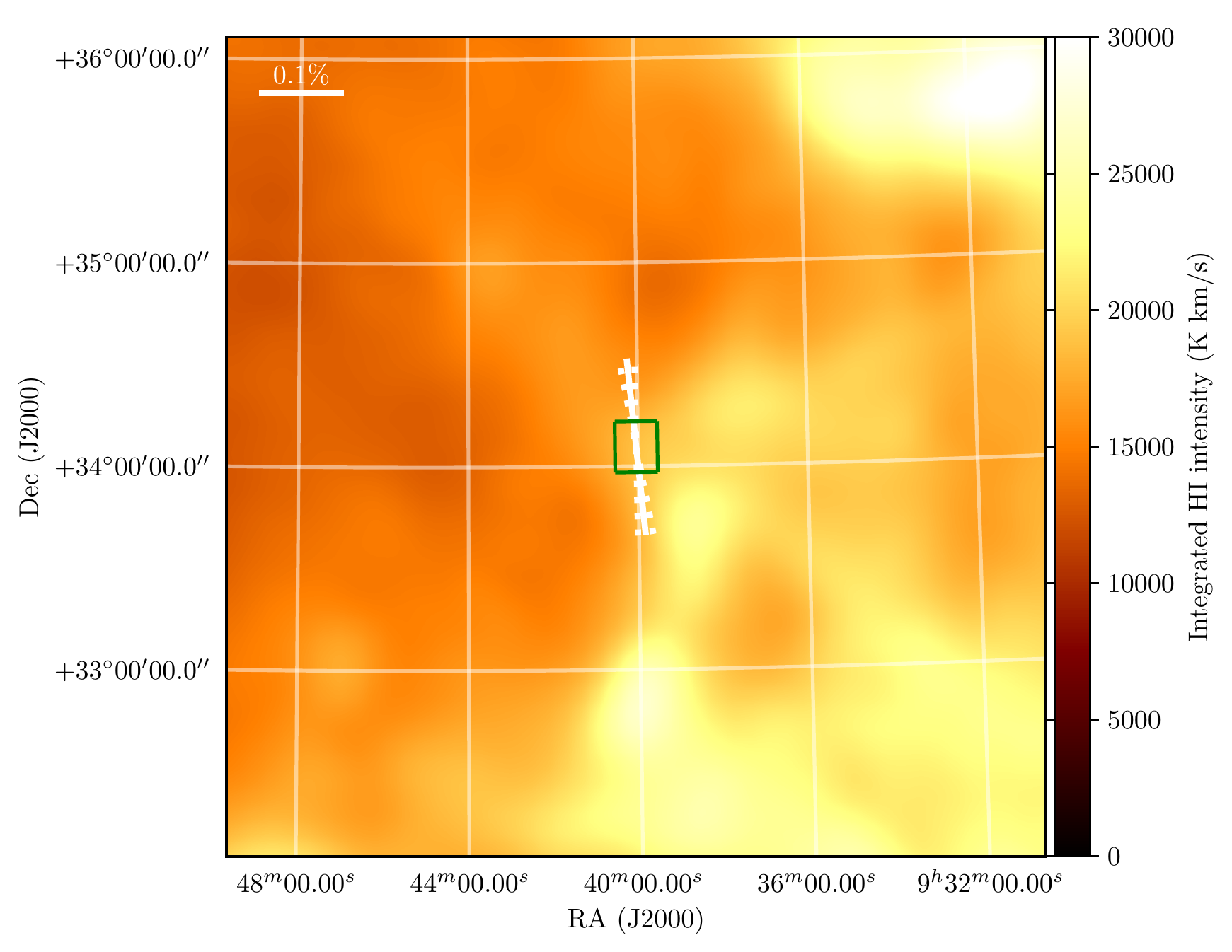}
    \end{minipage}
    \caption{As in Fig. \ref{fig:hi_data_dp2}, but for the two components in the spectrum of DP3. Top panel: velocity range centered on -13 km/s. Bottom panel: velocity range centered on -1 km/s. }
    \label{fig:hi_data_dp3}
  \end{figure}

    Inspecting the spectrum of DP3, we distinguish two prominent peaks at velocities that imply that the gas is local. These two intense components likely account for the main contribution to the polarization we observe. In the spectrum of DP2, there is a single component that dominates the signal, and it is also local. The marginally detected polarization of DP2 can be attributed mostly to this component. Meanwhile, the non-detection of polarization in DP1 could be the result of the much lower total intensity of the two components seen in the spectrum. Another alternative is that the component at high (absolute) velocities, which would be the main contributor to the column, is at a distance such that the bulk of the stars in our sample are foreground to it. Finally, there is also the possibility that the low polarization is a result of depolarization occurring along the line-of-sight to DP1, due to the presence of these two components. 
    
    By integrating over (approximately) the width of each velocity component, we can inspect the local morphology of the HI emission on the plane of the sky. Recent works have found the structure of the HI gas to be well correlated with the orientation of the magnetic field (using starlight polarization and HI data, \citet{clark}; or dust thermal emission and its polarization, \citet{planck_magnetic}). We investigate the relation of HI structure to the mean polarization in our regions.
    
    Figures \ref{fig:hi_data_dp2} and \ref{fig:hi_data_dp3} show maps of the HI intensity integrated over the velocity range [$v-\Delta v$, $v+ \Delta v$], where $\Delta v$ has been set equal to $6\rm \, km/s$ and $v$ is the velocity at which the spectra peak (including the secondary peak of DP3 at -1 $\rm km/s$). The maps are centered on DP2 (Fig. \ref{fig:hi_data_dp2}) and DP3 (Fig. \ref{fig:hi_data_dp3}). 
    DP1 is not presented because its $p_{mean}$ is consistent with zero.
    For each region, the black (or white) solid line segment forms an angle $\chi_{mean}$ with respect to the north (increasing towards the east according to the IAU convention for the EVPA) and shows the mean polarization orientation. Its length is proportional to $p_{mean}$ (Table \ref{table:stokes_weighted}). The dotted lines around the solid segment indicate the error in the mean polarization angle and the green boxes mark the surveyed regions. A polarization segment indicating the polarization scale is plotted in the top left of each image.
  
Inspecting Fig. \ref{fig:hi_data_dp2}, we find that the single component of the spectrum in DP2 shows a filamentary morphology. The orientation of the observed structure is not aligned with that of the mean polarization, at the resolution of 10$\arcmin$. The HI emission around DP3 is much more complex. The integrated intensity of the velocity component seen at -13 km/s is shown in the top panel of Fig. \ref{fig:hi_data_dp3}. DP3 appears to fall on the edge of a low-aspect-ratio (`blobby') structure, whose mean orientation is not clear. There does seem to be an asymmetry of the emission indicating an orientation from the south-east towards the north-west. The mean EVPA does not coincide with the axis of the asymmetry. The second velocity component, centered around -1 km/s, does not allow for a clear determination of a mean orientation in the HI morphology (bottom panel, Fig. \ref{fig:hi_data_dp3}). At higher resolution, such a comparison may be facilitated; for example, if the high-aspect-ratio structures identified by \citet{clark} in GALFA-HI data (4$\arcmin$) are present in this region. Finally, it is possible that the polarization of stars in DP3 is affected by both velocity components, meaning that an alignment of the polarization in each component with the corresponding HI structures cannot be excluded. 

\subsection{Comparison of $p_{mean}$ with E(B-V)}
    \label{sec:pav}
    
     Measurements of interstellar $p$ are bound by an upper envelope in the $p-E(B-V)$ plane. The first work to define such an envelope empirically was that of \citet{hiltner1956}, using observations of 1259 O and B stars with E(B-V) $\geq 0.13$ mag (we convert his presented total extinctions ($A_V$) to reddenings assuming a ratio of total-to-selective extinction $R_V = A_V/E(B-V) = 3.1$). He found that the majority of measurements are bound by the relation
    \begin{equation}
    p_{max}\approx 9 \, \rm{E(B-V) (\%/mag)},
    \label{eq:p_av}
    \end{equation}
    where we have used $R_V = 3.1$ and the conversion from $p$ in magnitudes to the fractional polarization (\cite{whittet}). Following works confirmed the existence of this envelope \citep[][E(B-V) $> 0.1$ mag]{serkowski1975}. In a much later work, \citet{fosalba} used stars with $\rm 0.01 \, mag < E(B-V) < 1$ mag to fit an expression for the mean $p$ as a function of E(B-V):
    \begin{equation} 
    \left\langle p \right\rangle = \rm 3.5*E(B-V)^{0.8} (\%/mag).
    \label{eq:fosalba}
    \end{equation}
We note that $\left\langle p \right\rangle$ is the mean $p$ of a sample of stars within a given range in E(B-V), and is therefore distinct from our determination of the mean fractional linear polarization within the DPs ($p_{mean}$).

    We wish to compare the $p_{mean}$ found in Sect. \ref{sec:results} to relations (\ref{eq:p_av}) and (\ref{eq:fosalba}) above. In order to make a fair comparison, we must take into account that the observed $p_{mean}$ is a biased estimator of the true $p_{mean}$ (e.g., \cite{simmons_stewart}). We calculate the debiased $p_{mean}$ using the formula of \citet{vaillancourt}: $\hat{p}_{mean} = \sqrt{p^2_{mean} - \sigma^2_{p,mean}}$, for DP2 and DP3 where $p_{mean} > 3 \sigma_{p,mean}$. The debiased estimate of $p_{mean}$ in DP1 is zero.
    
    A direct estimate of each star's reddening cannot be obtained with the available information on our sample. As an estimate of the mean reddening, we use the total E(B-V) from the LHD map towards each DP. This most likely overestimates the reddening for the stars that are not tracing the full line-of-sight. As there are systematic uncertainties associated with the conversion from HI column density to E(B-V), we adopt the following approach to estimate this uncertainty for the DPs. 
    \cite{schlegel} derived LHD E(B-V) from the relation of dust-emission-based E(B-V) to HI column, NHI (Fig. 1, right, in LHD). For a given NHI, there is a range of E(B-V) values
    observed, likely resulting from the combination of different effects (including variations in the gas-to-dust ratio and effects of cosmic infrared background fluctuations etc.). 
    For each DP, we have found the NHI and the spread of the corresponding E(B-V) $\pm$ 0.001 values (from the entire LHD sky footprint). We assign this spread as the error to the LHD E(B-V) in each DP.  
    
\begin{figure}
      \centering
      \includegraphics[width=\hsize]{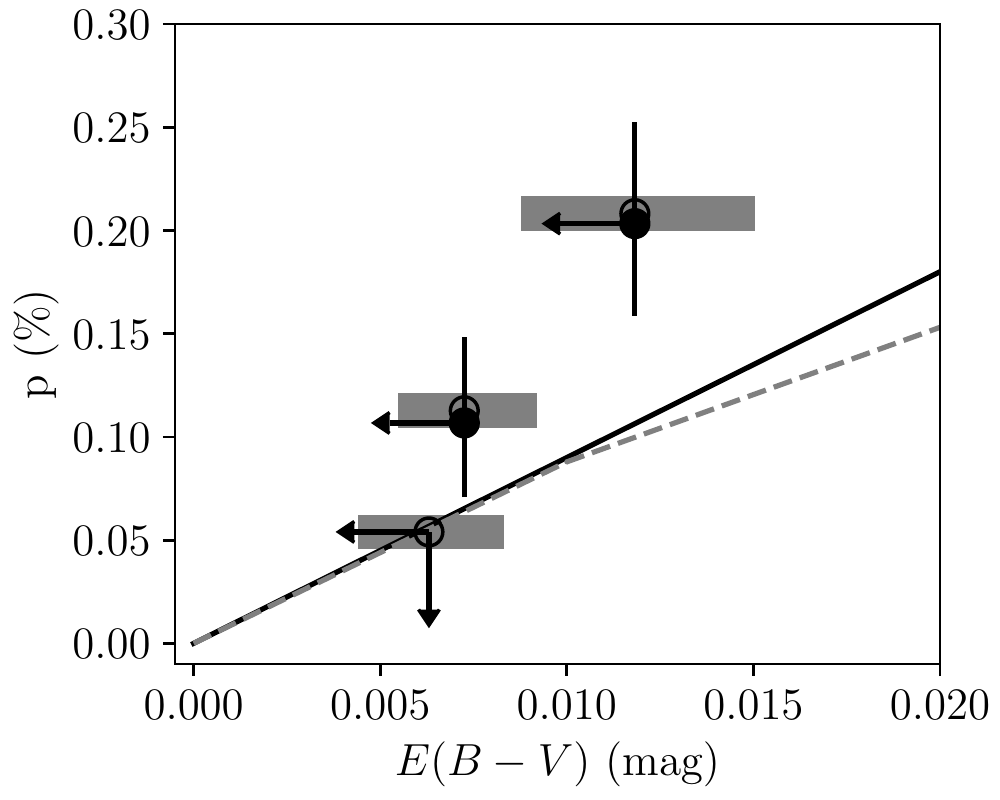}
      \caption{Fractional linear polarization vs. reddening. Open circles show the $p_{mean}$ of the DPs, while filled circles show the debiased $\hat{p}_{mean}$ (Table \ref{table:stokes_weighted}). The E(B-V) is found from the LHD map as explained in the text. The solid black line indicates the upper envelope from Eq. \ref{eq:p_av} \citep{hiltner1956}. The expression for the mean $p$, Eq. \ref{eq:fosalba} \citep{fosalba}, is traced by the dashed gray line. The gray regions indicate the interval in which E(B-V) values lie within $99\%$ confidence.}
      \label{fig:p_av_plot}
    \end{figure}

    In Figure \ref{fig:p_av_plot}, we compare our measurements to relations (\ref{eq:p_av}) and (\ref{eq:fosalba}). The figure shows both $p_{mean}$ and the debiased $\hat{p}_{mean}$ as open and filled circles, respectively. We find that both of our significant detections lie higher than both the $\left\langle p \right\rangle$ and $p_{max}$ curves. The $\hat{p}_{mean}$ of DP2 lies 1.1 $\sigma$ from both relations. The $\hat{p}_{mean}$ of DP3 lies 2.2 $\sigma$ from relation (\ref{eq:p_av}) and 2.3 $\sigma$ from relation (\ref{eq:fosalba}). Since our estimate of E(B-V) is an upper limit, it is likely that the points will be shifted towards the left, and the inconsistency with the $p_{max}$ envelope will be augmented.
        
    Polarization measurements at higher extinctions have consistently shown agreement with relation (\ref{eq:p_av}). Throughout the literature, only a small fraction of stellar $p$ measurements lie above the relation (e.g., Fig. 1(a) from \citet{andersson} and Fig. 15 in \citet{panop2015}). At E(B-V) $>$ 0.03 mag, the measurements of \citet{santos} are also largely consistent with this upper envelope. This is the first comparison with the aforementioned relations for such low extinctions.
    
    Interestingly, the relation for the upper envelope (\ref{eq:p_av}) coincides with that for $\left\langle p \right\rangle$ for E(B-V) $< 0.01$ mag. This implies that at least one of the two relations will lead to an erroneous estimate of $p$ for a given E(B-V). Both relations have been calculated at higher extinctions and have been extrapolated to these low E(B-V). However, in the case of the $p_{max}$ curve, the extrapolation has been made from E(B-V) that are an order of magnitude higher than the ones studied here. The \cite{fosalba} dataset used points down to E(B-V) $\sim 0.01$ mag and is therefore more reliable for these low extinctions. 
    
    This result implies that $p_{max}$ has previously been underestimated by an unknown factor, which may have implications on existing dust models \citep[e.g.,][]{planckxxi}. Even at extinctions of 0.1 mag, significant depolarization exists along the line-of-sight. If we consider a model of the 3D dust distribution as discrete polarizing `screens' (or clouds), our findings would suggest that the number of `screens' along the line-of-sight is much smaller than that at higher extinctions. In fact, the HI emission spectra of the DP fields in Sect. \ref{sec:hi} would suggest that 1-2 distinct components exist at these sightlines. We reserve a more detailed investigation of the polarizing efficiency of the low-dust-extinction sky for a future work.

\section{Discussion}
\label{sec:discussion}
 
    We set out to perform our stellar polarization mini-surveys in order to identify requirements for future optopolarimetric experiments targeting the high-latitude sky. With our setup of $\sim$ 20 stars in $\sim 0.05$ square degrees, and a systematic uncertainty of 0.1\% in $p$, we find the vast majority of individual stellar measurements to be non-detections. However, our strategy enables us to measure the mean fractional linear polarization within our target regions, with good enough precision so as to obtain significant detections in two out of three regions. With our systematic uncertainty and survey depth $R \leqslant 16$ mag, we obtain a marginal detection of $p_{mean}$ in DP2 ($\sim 0.11\% \pm 0.04 \%$, Table \ref{table:stokes_weighted}, or $p_{mean}/\sigma_{p,mean} \sim 3$). The level of the signal is low, but recoverable.  

Our survey strategy is qualitatively different from that of existing works at high latitudes, which have selected bright stars sampling sightlines that are distant on the plane of the sky. On the other hand, we have targeted a flux-limited sample in each of three very small areas of the sky. It is of interest to compare the results of the two different approaches. 

The vast majority of existing high-latitude measurements belong to the surveys of \cite{berdyugin2001_etc}, \cite{berdyugin2001}, \cite{berdyugin2002} and \cite{berdyugin2014}. These works have cataloged the polarization of over 2800 bright stars ($V\leqslant 13$ mag) with known distances (from the Hipparcos mission). Their sample is located at Galactic latitudes $b > 30^\circ, b < -60^\circ$. The bulk of stellar $p$ from the aforementioned catalogs are clustered around 0.1\%. We find that the $p_{mean}$ measured in this work are within the range of observed polarizations of the Berdyugin sample.

\section{Conclusions}
\label{sec:conclusions}
  
  In this work we performed stellar optical polarization surveys in three $\sim 15\arcmin \times 15\arcmin$ regions of the high-Galactic latitude sky. Our aim was to determine the level of interstellar fractional linear polarization ($p$) that can be recovered in regions of very low dust emission using the RoboPol polarimeter. Our surveys are photometrically complete down to R $\sim 16$ mag, and are therefore unique in depth for the high-Galactic-latitude sky.  Our findings can be summarized as follows.
    
     The determination of the systematic uncertainty is critical for our goal of measuring interstellar $p$ in  regions of very low dust emission. We have taken care to provide a reliable estimate of the uncertainty of our instrument for every observing run. We find the uncertainty to be at the level of 0.1\% in $p$ (in the focal plane mask where our observations were conducted) (Sect. \ref{sec:reduction}).
     Furthermore, we have provided a measurement of the intrinsic polarization of the standard star BD+33.2642 in the R-band (Appendix).
     
     We have detected significant $p$ for only seven stars in our samples. Most of them are outliers in the Stokes $q - u$ plane when compared to the location of the bulk of the measurements, but we could not identify indications of variability for these (or any other) stars in our sample (Sect. \ref{sec:results}).
     Even though most measurements have yielded only upper limits on $p$, we are able to locate the mean interstellar $p$ with high significance in the region with highest dust content (reddening), DP3, at the level of $p_{mean}=(0.208 \pm 0.044)\%$. In the region with intermediate reddening, DP2, we have obtained a marginal detection at 
 $p_{mean}=(0.113 \pm 0.036) \%$. In the region with least dust content, DP1, the $p_{mean}=(0.054 \pm 0.038) \%$, has not been significantly detected (Sect. \ref{sec:results}). 
    In addition, we have estimated the intrinsic spread of the Stokes $q$ and $u$ distributions, under the assumption that the distributions are Gaussian and have equal spread. We place upper limits on the intrinsic spread of: 0.187$(\%)$ for DP2 and 0.127$(\%)$ for DP3 (Sect. \ref{sec:results}).
   Through statistical tests we have assessed the confidence of the measured $p_{mean}$ and have found that the signal in DP2 and DP3 cannot be produced by uncertainties in our analysis (Sect. \ref{sec:stattests}).
   Using HI line emission from the EBHIS survey we identify two dominant components in DP3, one in DP2 and two much fainter components in DP1 (Sect. \ref{sec:hi}). The morphology of the component in DP2 is filamentary, and is not aligned with the mean EVPA at the resolution of 10$\arcmin$. In DP3 the morphology of both components is more complex and cannot be easily compared to the mean polarization angle (Sect. \ref{sec:hi}).

\begin{acknowledgements}
  
  We would like to thank the reviewer for insightful comments. In addition, we thank K. Kovlakas, T. Kougentakis, N. Mandarakas, G. Paterakis and A. Steiakaki for technical support and as well as N. D. Kylafis, H. Brandon, I. K. Wehus and D. Lenz for their helpful comments on the paper. RS would like to thank Dr K. Christidis for fruitful discussions. GVP acknowledges support from the National Science Foundation, under grant number AST-1611547.
  This project has received funding from the European Research Council (ERC) under the European Union's Horizon 2020 research and innovation programme under grant agreement No 771282.
  This publication makes use of VOSA, developed under the Spanish Virtual Observatory project supported from the Spanish MICINN through grant AyA2011-24052. This work has made use of data from the European Space Agency (ESA) mission {\it Gaia} (\url{https://www.cosmos.esa.int/gaia}), processed by the {\it Gaia} Data Processing and Analysis Consortium (DPAC, \url{https://www.cosmos.esa.int/web/gaia/dpac/consortium}). Funding for the DPAC has been provided by national institutions, in particular the institutions participating in the {\it Gaia} Multilateral Agreement.
      
\end{acknowledgements}


\bibliographystyle{aa}

\begin{thebibliography}{9}
 \bibitem[{{Andersson} {et~al.}(2015){Andersson}, {Lazarian}, \&
  {Vaillancourt}}]{andersson}
{Andersson}, B.-G., {Lazarian}, A., \& {Vaillancourt}, J.~E. 2015, \araa, 53,
  501

\bibitem[{{Angel} \& {Stockman}(1980)}]{angel}
{Angel}, J.~R.~P. \& {Stockman}, H.~S. 1980, \araa, 18, 321

\bibitem[{{Appenzeller}(1968)}]{appenzeller}
{Appenzeller}, I. 1968, \apj, 151, 907

\bibitem[{{Bailer-Jones} {et~al.}(2018){Bailer-Jones}, {Rybizki}, {Fouesneau},
  {Mantelet}, \& {Andrae}}]{bailer_jones}
{Bailer-Jones}, C.~A.~L., {Rybizki}, J., {Fouesneau}, M., {Mantelet}, G., \&
  {Andrae}, R. 2018, ArXiv e-prints [\eprint[arXiv]{1804.10121}]

\bibitem[{{Bailey} {et~al.}(2010){Bailey}, {Lucas}, \& {Hough}}]{bailey}
{Bailey}, J., {Lucas}, P.~W., \& {Hough}, J.~H. 2010, \mnras, 405, 2570

\bibitem[{{Bayo} {et~al.}(2008){Bayo}, {Rodrigo}, {Barrado Y Navascu{\'e}s},
  {Solano}, {Guti{\'e}rrez}, {Morales-Calder{\'o}n}, \& {Allard}}]{bayo}
{Bayo}, A., {Rodrigo}, C., {Barrado Y Navascu{\'e}s}, D., {et~al.} 2008, \aap,
  492, 277

\bibitem[{{Berdyugin} {et~al.}(2014){Berdyugin}, {Piirola}, \&
  {Teerikorpi}}]{berdyugin2014}
{Berdyugin}, A., {Piirola}, V., \& {Teerikorpi}, P. 2014, \aap, 561, A24

\bibitem[{{Berdyugin} \& {Teerikorpi}(2001)}]{berdyugin2001}
{Berdyugin}, A. \& {Teerikorpi}, P. 2001, \aap, 368, 635

\bibitem[{{Berdyugin} \& {Teerikorpi}(2002)}]{berdyugin2002}
{Berdyugin}, A. \& {Teerikorpi}, P. 2002, \aap, 384, 1050

\bibitem[{{Berdyugin} {et~al.}(2001){Berdyugin}, {Teerikorpi}, {Haikala},
  {Hanski}, {Knude}, \& {Markkanen}}]{berdyugin2001_etc}
{Berdyugin}, A., {Teerikorpi}, P., {Haikala}, L., {et~al.} 2001, \aap, 372, 276

\bibitem[{{BICEP2/Keck Collaboration} {et~al.}(2015){BICEP2/Keck
  Collaboration}, {Planck Collaboration}, {Ade}, {Aghanim}, {Ahmed}, {Aikin},
  {Alexander}, {Arnaud}, {Aumont}, {Baccigalupi}, \& et~al.}]{bicep}
{BICEP2/Keck Collaboration}, {Planck Collaboration}, {Ade}, P.~A.~R., {et~al.}
  2015, Physical Review Letters, 114, 101301

\bibitem[{{Clark} {et~al.}(2014){Clark}, {Peek}, \& {Putman}}]{clark}
{Clark}, S.~E., {Peek}, J.~E.~G., \& {Putman}, M.~E. 2014, \apj, 789, 82

\bibitem[{{Cudlip} {et~al.}(1982){Cudlip}, {Furniss}, {King}, \&
  {Jennings}}]{cudlip}
{Cudlip}, W., {Furniss}, I., {King}, K.~J., \& {Jennings}, R.~E. 1982, \mnras,
  200, 1169

\bibitem[{{Czekaj} {et~al.}(2014){Czekaj}, {Robin}, {Figueras}, {Luri}, \&
  {Haywood}}]{besancon}
{Czekaj}, M.~A., {Robin}, A.~C., {Figueras}, F., {Luri}, X., \& {Haywood}, M.
  2014, \aap, 564, A102

\bibitem[{{Davis} \& {Greenstein}(1951)}]{davisgreenstein}
{Davis}, Jr., L. \& {Greenstein}, J.~L. 1951, \apj, 114, 206

\bibitem[{{Drake} {et~al.}(2009){Drake}, {Djorgovski}, {Mahabal}, {Beshore},
  {Larson}, {Graham}, {Williams}, {Christensen}, {Catelan}, {Boattini},
  {Gibbs}, {Hill}, \& {Kowalski}}]{drake}
{Drake}, A.~J., {Djorgovski}, S.~G., {Mahabal}, A., {et~al.} 2009, \apj, 696,
  870

\bibitem[{{Fosalba} {et~al.}(2002){Fosalba}, {Lazarian}, {Prunet}, \&
  {Tauber}}]{fosalba}
{Fosalba}, P., {Lazarian}, A., {Prunet}, S., \& {Tauber}, J.~A. 2002, \apj,
  564, 762

\bibitem[{{Gaia Collaboration} {et~al.}(2018){Gaia Collaboration}, {Brown},
  {Vallenari}, {Prusti}, {de Bruijne}, {Babusiaux}, \& {Bailer-Jones}}]{gaia}
{Gaia Collaboration}, {Brown}, A.~G.~A., {Vallenari}, A., {et~al.} 2018, ArXiv
  e-prints [\eprint[arXiv]{1804.09365}]

\bibitem[{{Green} {et~al.}(2018){Green}, {Schlafly}, {Finkbeiner}, {Rix},
  {Martin}, {Burgett}, {Draper}, {Flewelling}, {Hodapp}, {Kaiser}, {Kudritzki},
  {Magnier}, {Metcalfe}, {Tonry}, {Wainscoat}, \& {Waters}}]{green_map}
{Green}, G.~M., {Schlafly}, E.~F., {Finkbeiner}, D., {et~al.} 2018, ArXiv
  e-prints [\eprint[arXiv]{1801.03555}]

\bibitem[{{Hall}(1949)}]{hall1949}
{Hall}, J.~S. 1949, Science, 109, 166

\bibitem[{{Heiles}(2000)}]{heiles}
{Heiles}, C. 2000, \aj, 119, 923

\bibitem[{{HI4PI Collaboration} {et~al.}(2016){HI4PI Collaboration}, {Ben
  Bekhti}, {Fl{\"o}er}, {Keller}, {Kerp}, {Lenz}, {Winkel}, {Bailin},
  {Calabretta}, {Dedes}, {Ford}, {Gibson}, {Haud}, {Janowiecki}, {Kalberla},
  {Lockman}, {McClure-Griffiths}, {Murphy}, {Nakanishi}, {Pisano}, \&
  {Staveley-Smith}}]{hi4pi}
{HI4PI Collaboration}, {Ben Bekhti}, N., {Fl{\"o}er}, L., {et~al.} 2016, \aap,
  594, A116

\bibitem[{{Hiltner}(1949)}]{hiltner2}
{Hiltner}, W.~A. 1949, \apj, 109, 471

\bibitem[{{Hiltner}(1956)}]{hiltner1956}
{Hiltner}, W.~A. 1956, \apjs, 2, 389

\bibitem[{{Jester} {et~al.}(2005){Jester}, {Schneider}, {Richards}, {Green},
  {Schmidt}, {Hall}, {Strauss}, {Vanden Berk}, {Stoughton}, {Gunn},
  {Brinkmann}, {Kent}, {Smith}, {Tucker}, \& {Yanny}}]{sebastian}
{Jester}, S., {Schneider}, D.~P., {Richards}, G.~T., {et~al.} 2005, \aj, 130,
  873

\bibitem[{{Kamionkowski} {et~al.}(1997{\natexlab{a}}){Kamionkowski},
  {Kosowsky}, \& {Stebbins}}]{kamionkowski1}
{Kamionkowski}, M., {Kosowsky}, A., \& {Stebbins}, A. 1997{\natexlab{a}},
  Physical Review Letters, 78, 2058

\bibitem[{{Kamionkowski} {et~al.}(1997{\natexlab{b}}){Kamionkowski},
  {Kosowsky}, \& {Stebbins}}]{kamionkowski2}
{Kamionkowski}, M., {Kosowsky}, A., \& {Stebbins}, A. 1997{\natexlab{b}}, \prd,
  55, 7368

\bibitem[{{King} {et~al.}(2014){King}, {Blinov}, {Ramaprakash}, {Myserlis},
  {Angelakis}, {Balokovi{\'c}}, {Feiler}, {Fuhrmann}, {Hovatta}, {Khodade},
  {Kougentakis}, {Kylafis}, {Kus}, {Modi}, {Paleologou}, {Panopoulou},
  {Papadakis}, {Papamastorakis}, {Paterakis}, {Pavlidou}, {Pazderska},
  {Pazderski}, {Pearson}, {Rajarshi}, {Readhead}, {Reig}, {Steiakaki},
  {Tassis}, \& {Zensus}}]{king}
{King}, O.~G., {Blinov}, D., {Ramaprakash}, A.~N., {et~al.} 2014, \mnras, 442,
  1706

\bibitem[{{Lenz} {et~al.}(2017){Lenz}, {Hensley}, \& {Dor{\'e}}}]{lenz}
{Lenz}, D., {Hensley}, B.~S., \& {Dor{\'e}}, O. 2017, \apj, 846, 38

\bibitem[{{Leroy}(1993)}]{leroy}
{Leroy}, J.~L. 1993, \aaps, 101, 551

\bibitem[{{Luri} {et~al.}(2018){Luri}, {Brown}, {Sarro}, {Arenou},
  {Bailer-Jones}, {Castro-Ginard}, {de Bruijne}, {Prusti}, {Babusiaux}, \&
  {Delgado}}]{luri_brown}
{Luri}, X., {Brown}, A.~G.~A., {Sarro}, L.~M., {et~al.} 2018, ArXiv e-prints
  [\eprint[arXiv]{1804.09376}]

\bibitem[{{Monet} {et~al.}(2003){Monet}, {Levine}, {Canzian}, {Ables}, {Bird},
  {Dahn}, {Guetter}, {Harris}, {Henden}, {Leggett}, {Levison}, {Luginbuhl},
  {Martini}, {Monet}, {Munn}, {Pier}, {Rhodes}, {Riepe}, {Sell}, {Stone},
  {Vrba}, {Walker}, {Westerhout}, {Brucato}, {Reid}, {Schoening}, {Hartley},
  {Read}, \& {Tritton}}]{usnob}
{Monet}, D.~G., {Levine}, S.~E., {Canzian}, B., {et~al.} 2003, \aj, 125, 984

\bibitem[{{Panopoulou} {et~al.}(2015){Panopoulou}, {Tassis}, {Blinov},
  {Pavlidou}, {King}, {Paleologou}, {Ramaprakash}, {Angelakis},
  {Balokovi{\'c}}, {Das}, {Feiler}, {Hovatta}, {Khodade}, {Kiehlmann}, {Kus},
  {Kylafis}, {Liodakis}, {Mahabal}, {Modi}, {Myserlis}, {Papadakis},
  {Papamastorakis}, {Pazderska}, {Pazderski}, {Pearson}, {Rajarshi},
  {Readhead}, {Reig}, \& {Zensus}}]{panop2015}
{Panopoulou}, G., {Tassis}, K., {Blinov}, D., {et~al.} 2015, \mnras, 452, 715

\bibitem[{{Piirola}(1977)}]{piirola}
{Piirola}, V. 1977, \aaps, 30, 213

\bibitem[{{Planck Collaboration} {et~al.}(2014){Planck Collaboration},
  {Abergel}, {Ade}, {Aghanim}, {Alves}, {Aniano}, {Armitage-Caplan}, {Arnaud},
  {Ashdown}, {Atrio-Barandela}, \& et~al.}]{dust_map}
{Planck Collaboration}, {Abergel}, A., {Ade}, P.~A.~R., {et~al.} 2014, \aap,
  571, A11

\bibitem[{{Planck Collaboration} {et~al.}(2016{\natexlab{a}}){Planck
  Collaboration}, {Adam}, {Ade}, {Aghanim}, {Alves}, {Arnaud}, {Arzoumanian},
  {Ashdown}, {Aumont}, {Baccigalupi}, \& et~al.}]{planck_magnetic}
{Planck Collaboration}, {Adam}, R., {Ade}, P.~A.~R., {et~al.}
  2016{\natexlab{a}}, \aap, 586, A135

\bibitem[{{Planck Collaboration} {et~al.}(2015){Planck Collaboration}, {Ade},
  {Aghanim}, {Alina}, {Aniano}, {Armitage-Caplan}, {Arnaud}, {Ashdown},
  {Atrio-Barandela}, {Aumont}, {Baccigalupi}, {Banday}, {Barreiro}, {Battaner},
  {Beichman}, {Benabed}, {Benoit-L{\'e}vy}, {Bernard}, {Bersanelli},
  {Bielewicz}, {Bock}, {Bond}, {Borrill}, {Bouchet}, {Boulanger}, {Burigana},
  {Cardoso}, {Catalano}, {Chamballu}, {Chary}, {Chiang}, {Christensen},
  {Colombi}, {Colombo}, {Combet}, {Couchot}, {Coulais}, {Crill}, {Curto},
  {Cuttaia}, {Danese}, {Davies}, {Davis}, {de Bernardis}, {de Rosa}, {de
  Zotti}, {Delabrouille}, {D{\'e}sert}, {Dickinson}, {Diego}, {Donzelli},
  {Dor{\'e}}, {Douspis}, {Dunkley}, {Dupac}, {Efstathiou}, {En{\ss}lin},
  {Eriksen}, {Falgarone}, {Fanciullo}, {Finelli}, {Forni}, {Frailis},
  {Fraisse}, {Franceschi}, {Galeotta}, {Ganga}, {Ghosh}, {Giard},
  {Giraud-H{\'e}raud}, {Gonz{\'a}lez-Nuevo}, {G{\'o}rski}, {Gregorio},
  {Gruppuso}, {Guillet}, {Hansen}, {Harrison}, {Helou},
  {Hern{\'a}ndez-Monteagudo}, {Hildebrandt}, {Hivon}, {Hobson}, {Holmes},
  {Hornstrup}, {Huffenberger}, {Jaffe}, {Jaffe}, {Jones}, {Juvela},
  {Keih{\"a}nen}, {Keskitalo}, {Kisner}, {Kneissl}, {Knoche}, {Kunz},
  {Kurki-Suonio}, {Lagache}, {L{\"a}hteenm{\"a}ki}, {Lamarre}, {Lasenby},
  {Lawrence}, {Leonardi}, {Levrier}, {Liguori}, {Lilje}, {Linden-V{\o}rnle},
  {L{\'o}pez-Caniego}, {Lubin}, {Mac{\'{\i}}as-P{\'e}rez}, {Maffei},
  {Magalh{\~a}es}, {Maino}, {Mandolesi}, {Maris}, {Marshall}, {Martin},
  {Mart{\'{\i}}nez-Gonz{\'a}lez}, {Masi}, {Matarrese}, {Mazzotta},
  {Melchiorri}, {Mendes}, {Mennella}, {Migliaccio}, {Miville-Desch{\^e}nes},
  {Moneti}, {Montier}, {Morgante}, {Mortlock}, {Munshi}, {Murphy}, {Naselsky},
  {Nati}, {Natoli}, {Netterfield}, {Noviello}, {Novikov}, {Novikov},
  {Oxborrow}, {Pagano}, {Pajot}, {Paladini}, {Paoletti}, {Pasian}, {Perdereau},
  {Perotto}, {Perrotta}, {Piacentini}, {Piat}, {Pietrobon}, {Plaszczynski},
  {Poidevin}, {Pointecouteau}, {Polenta}, {Popa}, {Pratt}, {Prunet}, {Puget},
  {Rachen}, {Reach}, {Rebolo}, {Reinecke}, {Remazeilles}, {Renault},
  {Ricciardi}, {Riller}, {Ristorcelli}, {Rocha}, {Rosset}, {Roudier},
  {Rusholme}, {Sandri}, {Savini}, {Scott}, {Spencer}, {Stolyarov}, {Stompor},
  {Sudiwala}, {Sutton}, {Suur-Uski}, {Sygnet}, {Tauber}, {Terenzi},
  {Toffolatti}, {Tomasi}, {Tristram}, {Tucci}, {Umana}, {Valenziano},
  {Valiviita}, {Van Tent}, {Vielva}, {Villa}, {Wade}, {Wandelt}, \&
  {Zonca}}]{planckxxi}
{Planck Collaboration}, {Ade}, P.~A.~R., {Aghanim}, N., {et~al.} 2015, \aap,
  576, A106

\bibitem[{{Planck Collaboration} {et~al.}(2016{\natexlab{b}}){Planck
  Collaboration}, {Ade}, {Aghanim}, {Alves}, {Aniano}, {Arnaud}, {Ashdown},
  {Aumont}, {Baccigalupi}, {Banday}, {Barreiro}, {Bartolo}, {Battaner},
  {Benabed}, {Benoit-L{\'e}vy}, {Bernard}, {Bersanelli}, {Bielewicz},
  {Bonaldi}, {Bonavera}, {Bond}, {Borrill}, {Bouchet}, {Boulanger}, {Burigana},
  {Butler}, {Calabrese}, {Cardoso}, {Catalano}, {Chamballu}, {Chiang},
  {Christensen}, {Clements}, {Colombi}, {Colombo}, {Couchot}, {Crill}, {Curto},
  {Cuttaia}, {Danese}, {Davies}, {Davis}, {de Bernardis}, {de Rosa}, {de
  Zotti}, {Delabrouille}, {Dickinson}, {Diego}, {Dole}, {Donzelli}, {Dor{\'e}},
  {Douspis}, {Draine}, {Ducout}, {Dupac}, {Efstathiou}, {Elsner}, {En{\ss}lin},
  {Eriksen}, {Falgarone}, {Finelli}, {Forni}, {Frailis}, {Fraisse},
  {Franceschi}, {Frejsel}, {Galeotta}, {Galli}, {Ganga}, {Ghosh}, {Giard},
  {Gjerl{\o}w}, {Gonz{\'a}lez-Nuevo}, {G{\'o}rski}, {Gregorio}, {Gruppuso},
  {Guillet}, {Hansen}, {Hanson}, {Harrison}, {Henrot-Versill{\'e}},
  {Hern{\'a}ndez-Monteagudo}, {Herranz}, {Hildebrandt}, {Hivon}, {Holmes},
  {Hovest}, {Huffenberger}, {Hurier}, {Jaffe}, {Jaffe}, {Jones},
  {Keih{\"a}nen}, {Keskitalo}, {Kisner}, {Kneissl}, {Knoche}, {Kunz},
  {Kurki-Suonio}, {Lagache}, {Lamarre}, {Lasenby}, {Lattanzi}, {Lawrence},
  {Leonardi}, {Levrier}, {Liguori}, {Lilje}, {Linden-V{\o}rnle},
  {L{\'o}pez-Caniego}, {Lubin}, {Mac{\'{\i}}as-P{\'e}rez}, {Maffei}, {Maino},
  {Mandolesi}, {Maris}, {Marshall}, {Martin}, {Mart{\'{\i}}nez-Gonz{\'a}lez},
  {Masi}, {Matarrese}, {Mazzotta}, {Melchiorri}, {Mendes}, {Mennella},
  {Migliaccio}, {Miville-Desch{\^e}nes}, {Moneti}, {Montier}, {Morgante},
  {Mortlock}, {Munshi}, {Murphy}, {Naselsky}, {Natoli}, {N{\o}rgaard-Nielsen},
  {Novikov}, {Novikov}, {Oxborrow}, {Pagano}, {Pajot}, {Paladini}, {Paoletti},
  {Pasian}, {Perdereau}, {Perotto}, {Perrotta}, {Pettorino}, {Piacentini},
  {Piat}, {Plaszczynski}, {Pointecouteau}, {Polenta}, {Ponthieu}, {Popa},
  {Pratt}, {Prunet}, {Puget}, {Rachen}, {Reach}, {Rebolo}, {Reinecke},
  {Remazeilles}, {Renault}, {Ristorcelli}, {Rocha}, {Roudier},
  {Rubi{\~n}o-Mart{\'{\i}}n}, {Rusholme}, {Sandri}, {Santos}, {Scott},
  {Spencer}, {Stolyarov}, {Sudiwala}, {Sunyaev}, {Sutton}, {Suur-Uski},
  {Sygnet}, {Tauber}, {Terenzi}, {Toffolatti}, {Tomasi}, {Tristram}, {Tucci},
  {Umana}, {Valenziano}, {Valiviita}, {Van Tent}, {Vielva}, {Villa}, {Wade},
  {Wandelt}, {Wehus}, {Ysard}, {Yvon}, {Zacchei}, \& {Zonca}}]{planck_ebv}
{Planck Collaboration}, {Ade}, P.~A.~R., {Aghanim}, N., {et~al.}
  2016{\natexlab{b}}, \aap, 586, A132

\bibitem[{{Santos} {et~al.}(2011){Santos}, {Corradi}, \& {Reis}}]{santos}
{Santos}, F.~P., {Corradi}, W., \& {Reis}, W. 2011, \apj, 728, 104

\bibitem[{{Schlegel} {et~al.}(1998){Schlegel}, {Finkbeiner}, \&
  {Davis}}]{schlegel}
{Schlegel}, D.~J., {Finkbeiner}, D.~P., \& {Davis}, M. 1998, \apj, 500, 525

\bibitem[{{Schmidt} {et~al.}(1992){Schmidt}, {Elston}, \& {Lupie}}]{schmidt}
{Schmidt}, G.~D., {Elston}, R., \& {Lupie}, O.~L. 1992, \aj, 104, 1563

\bibitem[{{Schultz} \& {Wiemer}(1975)}]{schultz_wiemer}
{Schultz}, G.~V. \& {Wiemer}, W. 1975, \aap, 43, 133

\bibitem[{{Seljak}(1997)}]{seljak}
{Seljak}, U. 1997, \apj, 482, 6

\bibitem[{{Seljak} \& {Zaldarriaga}(1997)}]{seljak_zalda}
{Seljak}, U. \& {Zaldarriaga}, M. 1997, Physical Review Letters, 78, 2054

\bibitem[{{Serkowski} {et~al.}(1975){Serkowski}, {Mathewson}, \&
  {Ford}}]{serkowski1975}
{Serkowski}, K., {Mathewson}, D.~S., \& {Ford}, V.~L. 1975, \apj, 196, 261

\bibitem[{{Simmons} \& {Stewart}(1985)}]{simmons_stewart}
{Simmons}, J.~F.~L. \& {Stewart}, B.~G. 1985, \aap, 142, 100

\bibitem[{{Stein}(1966)}]{stein}
{Stein}, W. 1966, \apj, 144, 318

\bibitem[{{Tassis} \& {Pavlidou}(2015)}]{tassis_pavlidou}
{Tassis}, K. \& {Pavlidou}, V. 2015, \mnras, 451, L90

\bibitem[{{Tinbergen}(1982)}]{tinbergen}
{Tinbergen}, J. 1982, \aap, 105, 53

\bibitem[{{Turnshek} {et~al.}(1990){Turnshek}, {Bohlin}, {Williamson}, {Lupie},
  {Koornneef}, \& {Morgan}}]{turnshek}
{Turnshek}, D.~A., {Bohlin}, R.~C., {Williamson}, II, R.~L., {et~al.} 1990,
  \aj, 99, 1243

\bibitem[{{Vaillancourt}(2006)}]{vaillancourt}
{Vaillancourt}, J.~E. 2006, \pasp, 118, 1340

\bibitem[{{Venters} \& {Pavlidou}(2007)}]{venters}
{Venters}, T.~M. \& {Pavlidou}, V. 2007, \apj, 666, 128

\bibitem[{{Whittet}(1992)}]{whittet}
{Whittet}, D.~C.~B. 1992, {Dust in the galactic environment}

\bibitem[{{Winkel} {et~al.}(2016){Winkel}, {Kerp}, {Fl{\"o}er}, {Kalberla},
  {Ben Bekhti}, {Keller}, \& {Lenz}}]{winkel}
{Winkel}, B., {Kerp}, J., {Fl{\"o}er}, L., {et~al.} 2016, \aap, 585, A41

\bibitem[{{Zaldarriaga} \& {Seljak}(1997)}]{zaldarriaga1}
{Zaldarriaga}, M. \& {Seljak}, U. 1997, \prd, 55, 1830

\end{thebibliography}

\appendix
\section{Reference value of the polarization standard star BD+33.2642 in the R$-$band}
\label{sec:appendix}

During the course of the 2017 RoboPol observing season, a large sample of polarization standard stars was observed with high cadence. The wealth of data, combined with the stability of the instrument, enables us to characterize the properties of individual standard stars. 
We select the most well-sampled and stable stars from our set of calibrators (to minimize any possibility of intrinsic variability) and 
use them to derive the Stokes parameters of BD+33.2642 in the $R-$band.
Apart from the standard stars in Table \ref{table:standards_table}, we also include the star HD154892 \citep[$p = 0.05 \pm 0.03$, in the B-band,][]{turnshek} in this analysis.

\begin{figure}
   \centering
   \includegraphics[width=\hsize]{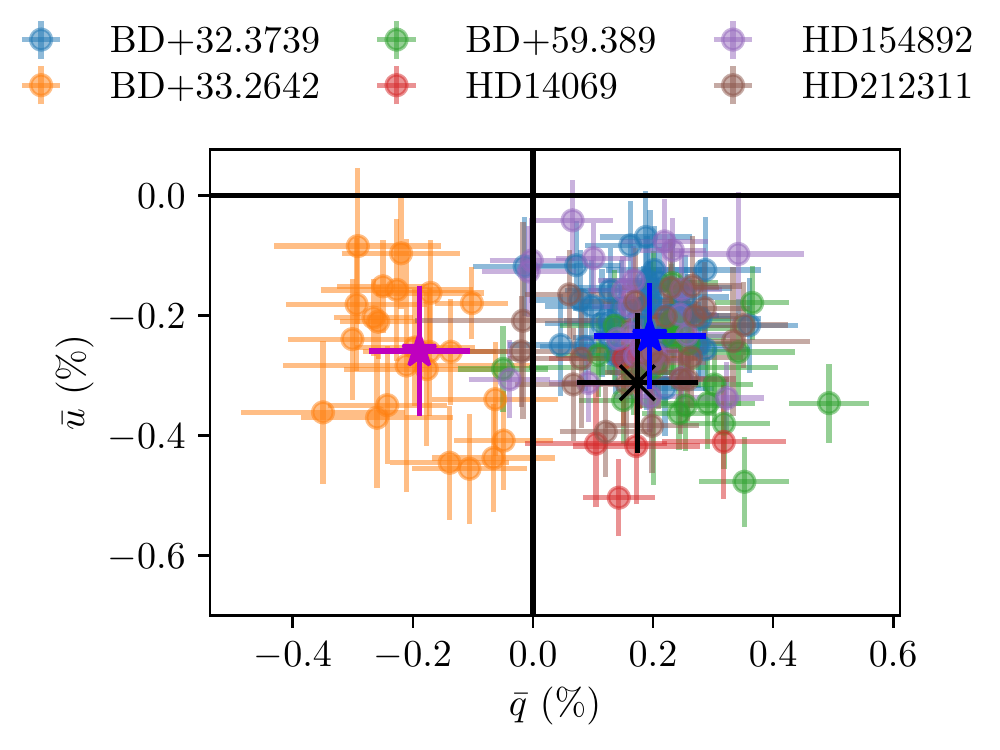}
      \caption{Measurements of the residual $q$,$u$ of standard stars after subtraction of their literature values (Table \ref{table:standards_table}) (circles: each color represents a different star). Blue star: weighted mean of $\bar{q},\bar{u}$ excluding BD+33.2642. Purple star: weighted mean of BD+33.2642 measurements corrected using the literature value in the $V-$band (table \ref{table:BD+332642}. The black `x' marks the weighted mean found for DP2 in Sect. \ref{sec:calibration}.
              }
         \label{fig:bdcompare}
   \end{figure}

Figure \ref{fig:bdcompare} shows the measurements of $\bar{q},\bar{u}$ (literature-subtracted measurements of standards, following Sect. \ref{sec:calibration}) for this set of stars observed throughout the 2017 season (May-November). All observations were conducted in the $R-$band and reduced as in Sect. \ref{sec:reduction}. 
For BD+33.2642, we use its literature value for the $V-$band (Table \ref{table:standards_table}). It is clear that this star exhibits an offset from the bulk of standards in the $\bar{q}$, $\bar{u}$ plane. Calculating the weighted mean of the $\bar{q}$ of BD+33.2642 (marked with a purple star in Fig. \ref{fig:bdcompare}), we find that it lies  $\sim 3 \sigma$ away from the weighted mean $\bar{q}$ of the other star measurements (blue star). We show the weighted mean from Table \ref{table:instrumental_error} for DP2 for comparison (black x). This was found using only standards from the nights when DP2 targets were observed (black x). The two zero-point estimates are within 1$\sigma$ of each other, demonstrating the stability of the instrument.

The only plausible explanation for the observed offset of BD+33.2642 is that its polarization properties in the R-band are different than those in the V-band. 
We calculate its `true' Stokes parameters in the $R-$band in the following way. We find the weighted mean $\left\langle\bar{q}\right\rangle,\left\langle\bar{u}\right\rangle$ of the measurements of all other standard stars. We also find the weighted mean of the measurements of BD+33.2642 (not corrected for the literature value) $\left\langle q_{B33}\right\rangle,\left\langle u_{B33}\right\rangle$. An estimate of the `true' Stokes parameters is: 
\begin{equation}
q_{true} = \left\langle q_{B33}\right\rangle - \left\langle\bar{q}\right\rangle,\
u_{true} = \left\langle u_{B33}\right\rangle - \left\langle\bar{u}\right\rangle 
,\end{equation}
and their uncertainty:
\begin{equation}
\sigma_{q,{true}} = \sqrt{\sigma_{\left\langle q\right\rangle,B33}^2 + \sigma_{\left\langle\bar{q}\right\rangle}^2}, \
\sigma_{u,{true}} = \sqrt{\sigma_{\left\langle u\right\rangle,B33}^2 + \sigma_{\left\langle\bar{u}\right\rangle}^2},
\end{equation}
where $\sigma_{\left\langle q\right\rangle,B33}$ and $\sigma_{\left\langle\bar{q}\right\rangle}$ (and similarly for $u$) are the weighted standard deviations of the values used for calculating $\left\langle q_{B33}\right\rangle$ and  $\left\langle\bar{q}\right\rangle $, respectively. As in Sect. \ref{sec:calibration}, we use the weighted standard deviation of measurements as an estimate for the error on the weighted mean, as a conservative choice. In this way, we take into account any possible intrinsic variability of the sources.

The resulting values for the polarization of BD+33.2642 are shown in Table \ref{table:BD+332642}. The star has been observed in previous years less times than in 2017. We have not used measurements from previous year for this determination. However, we have checked that repeating the same process with the data from seasons 2015 and 2016 does not produce `true' polarization parameters that are inconsistent with the determination presented here.

We note that in the analysis of Sect. \ref{sec:calibration}, the $R-$band polarization of BD+33.2642 is used only for the calibration of DP1, which was observed in 2015. Therefore, we are using completely independent measurements to derive its `true' polarization. 
  
       \begin{table}
      \caption{Polarization of BD+33.2642 in $V-$band (left) and $R-$band (right).}
      \label{table:BD+332642}
      \centering                          
      \begin{tabular}{c |c| c }        
        \hline\hline                 
    &\cite{schmidt}& This work\\
    \hline
        $q (\%)$ & 0.225&  -0.19 $\pm$ 0.15   \\
    $u (\%)$& 0.05 & 0.08 $\pm$ 0.14\\
    $p  (\%)$& 0.231$\pm$0.031& 0.20$\pm$ 0.15\\
    $\chi (^{o}) $& 12.67$^{\circ}$&  78$\pm$ 20 $^{\circ}$ \\ 
        \hline                                   
      \end{tabular}
     \end{table}

As a final remark, we find the polarization angle in the R-band to be significantly different from that in the V-band. This may indicate that the polarization properties of BD+33.2642 have changed since the original measurements of \cite{schmidt}. This calls for verification with the use of different instruments.

\end{document}